\documentclass[aps,twocolumn, amsmath,amssymb, pra]{revtex4}
\usepackage{graphicx}
\usepackage{epsfig}

\begin{document}

\title{Fermion- and spin-counting in strongly correlated systems in and out of thermal equilibrium}

\author{Sibylle Braungardt\(^1\), Mirta Rodr\' iguez\(^2\) , Aditi Sen(De)\(^3\), Ujjwal Sen\(^3\), Roy J. Glauber\(^4\), and Maciej
Lewenstein\(^{*,1}\)}
\affiliation{\(^1\)ICFO-Institut de Ci\`encies Fot\`oniques, Mediterranean Technology Park,
08860 Castelldefels (Barcelona), Spain\\
\(^2\)Instituto de Estructura de la Materia, CSIC, C/Serrano 121 28006 Madrid, Spain \\
\(^3\)Harish-Chandra Research
Institute, Chhatnag Road, Jhunsi, Allahabad 211 019, India\\
\(^4\)Lyman Laboratory, Physics Department, Harvard University, 02138 Cambridge, MA, U.S.A. \\
\(^*\)ICREA-Instituci\`o Catala de Ricerca i Estudis Avan{\c c}ats, 08010 Barcelona, Spain}

\begin{abstract}

Atom counting theory can be used to study the role of thermal
noise in quantum phase transitions and to monitor the
dynamics of a quantum system. We illustrate this for a strongly correlated fermionic
system, which is equivalent to an anisotropic quantum XY chain in a
transverse field, and can be realized with cold fermionic atoms in an optical
lattice. We analyze the counting statistics across the phase diagram in the presence of thermal fluctuations,
 and during its thermalization
when the system is coupled to a heat bath. At zero temperature,
the quantum phase transition is reflected in the cumulants of the
counting distribution. We find that the signatures of the
crossover remain visible at low temperature and are obscured with
increasing thermal fluctuations. We find that the same quantities may
be used to scan the dynamics during the thermalization of the system.
\end{abstract}

\maketitle
\def\bra#1{\langle#1|} \def\ket#1{|#1\rangle}
\def \av#1{\langle #1\rangle}
\def\com#1{{\tt [\hskip.5cm #1 \hskip.5cm ]}}

\section{Introduction}

In the past decade it became clear that the  most important
challenges of physics of ultracold atoms overlap essentially with
those of condensed matter physics, and concern  strongly
correlated quantum states of many body systems.  In fact,
ultracold fermionic and bosonic atoms in optical lattices mimic
strongly correlated systems, that can be perfectly described by
various Hubbard or spin  models
   with rich phase diagrams \cite{adp}.

   Amazingly, atomic physics
   may address questions concerning both static and dynamical
   properties of such systems. In the context of statics, the
   goal
   is to {\it quantum engineer}, i.e. to prepare, or reach interesting quantum phases or states,
   and then to detect their properties.  Many examples of such exotic
phases pertaining to quantum magnetism  based on super-exchange
interactions are now within experimental reach \cite{expe,expe2D}.
 Also, the signatures of itinerant ferromagnetism
in the absence of the lattice structure have been recently
reported for a system of spin 1/2 fermions \cite{expek}.

Despite the progress of experimental techniques, the preparation
and detection of quantum magnetism is always obscured by the
unavoidable noise and thermal effects. These are particularly
important in low dimensional systems, and, in particular, in  one dimension (1D)
where no long range order can exist at $T>0$. It is therefore
highly desirable to design detection methods that will allow to
observe  the signatures of strong correlations and quantum phase
transitions (QFT) at $T>0$. The first goal of this paper is to
demonstrate that atom counting may be used to detect signatures of
QFTs at $T>0$. To this aim we analyze a paradigmatic example of a
strongly correlated system: a  system of fermions in a
1D optical lattice.

Remarkable long time scales in ultracold atom experiments allow to
monitor the dynamics of the system directly. In the context of
dynamics, one goal is to observe the time evolution of the system
under some perturbation as the system approaches a stationary state. In
this context various fundamental questions can be addressed. For
instance, does the system, which can be very well regarded as
closed, thermalize after initial perturbation (sudden quench)
\cite{Sengupta,sensother,Muramatsu,Eisert,ortiz,Ciracbanuls}? What is the
difference between thermal and non thermal dynamics? What kinds of
interesting dynamical processes involving a coupling to a
specially designed heat bath can be realized? Can one realize
state engineering using open system dynamics \cite{env}? The
second goal of this paper is thus to study atom counting during
dynamic evolution. In particular, we compute the atom counting
distributions as a function of time when the analyzed 1D system of
fermions approaches the quantum Boltzmann-Gibbs thermal
equilibrium state at certain $T>0$. We show how the thermalization
process can be monitored by observing the cumulants of the
counting distribution. In principle, the methods allows thus to
distinguish thermal dynamics from   non-thermal one.

Counting of particles is one of the most important techniques of
characterization of quantum mechanical states of many body
systems. Photon counting, whose theory was developed in the
seminal works \cite{Glauber-seminal}, allows for the
full characterization of quantum light sources. More recently, the
counting statistics of electrons has been used to characterize
mesoscopic devices
\cite{electrons,bednorz,belzig,bbb,oppen,electrons2,schonhammer}.
In both mentioned cases, the particles considered are non-, or
practically noninteracting. In this paper, in contrast, we
consider strongly correlated atomic systems \cite{Sachdev}.
Counting statistics of atoms has been suggested as a technique to detect and distinguish various quantum phases of spin and
fermionic systems
\cite{demler,counting,maciejlewenstein,fermiones,bruder2}. Atom
counting can be realized in several manners (for early experiments,
see \cite{shimizu}). One method concerns metastable atoms, such as
metastable Helium \cite{Helium} - the atoms are released here from
the trap, so counting is preceded by essentially ballistic
expansion of the atomic wave functions. With the recent development
of high-resolution optical imaging systems, single atoms can be
detected with near-unit fidelity on individual sites of an optical
lattice \cite{expeg,expeb}. This makes available the counting
distributions of atoms {\it in situ} in the lattice. On the other
hand, spin counting techniques \cite{Sorensen} allow for the
measurement of the average and fluctuations of the spin number
also {\it in situ} in cold atomic samples. These techniques can be
extended to account for spatial resolution \cite{tobenat} and give
access to the Fourier components of the spin distribution
\cite{Troscilde}. With the help of superlattice configurations,
one may address the atoms locally, probing e.g. every second site
\cite{tobenat}. In this work we focus on {\it in situ} methods,
leaving the discussion  of the interplay of  atomic cloud
expansion and atom counting to a separate publication.

Despite the fact that in experimental
conditions noise (thermal or non-thermal) is always present, so
far atom counting has been mainly considered at zero temperature
and in the absence of non-thermal noise
\cite{demler,counting,fermiones,bruder2,maciejlewenstein},
 In particular, in Ref.
\cite{counting}, we have used atom counting theory to study a
system of fermions in a one-dimensional (1D) optical lattice. We
have shown that the critical behavior of the system, and in
particular formation of fermionic pairs,  is reflected in the
cumulants of the counting distribution. Here,  we consider the
counting distribution of the same fermionic system, but we now
take into account the effect of thermal noise, both by considering the effects of temperature when the system is in an equilibrium state and by that of thermalization when the system is attached to a model heat bath. Fermionic pair
breaking induced by thermal noise is clearly reflected in the
counting distribution function. We find also that the signatures
of the crossover between different phases
 remain visible at low
temperature, and we show how they fade out as the temperature
increases.

The paper is organized as follows. In Sec. \ref{sec-fermi}, we
provide a description of the fermion and spin system that we
consider.  In section \ref{sec_counting} we review the counting
theory for a fermionic system, and show how the counting
distribution can be obtained from a simple recursive formula.
Details of how to derive the counting distribution in terms of a
generating function are shown in the Appendix. In section
\ref{sec-counting_equi}, we study the counting statistics of the
system at thermal equilibrium at non-zero temperatures. First, in
subsection \ref{sec-counting_zero}, we present  the counting
distribution at zero temperature for reference.  Then, in
subsection \ref{sec-fert},
 we analyze how thermal noise affects  the atom number
 distributions, especially in the vicinity of  the quantum phase
 transition, or better to say cross-over.
In Sec. \ref{sec-counting_heat}, we calculate the atom number
distributions during a model thermalization process, in which the
system is coupled to a heat bath via the exchange of collective
quasi-particles. Such couplings, and the resulting open system
dynamics are  not strictly speaking local. In Sec. \ref{sec-local}
we analyze, however,  the nature of these couplings more closely,
and show that they can be well approximated by a physically
feasible model of local exchange of atoms between the system and
the reservoir. We summarize our results in Sec. \ref{sec-conc}.

\section{Fermi Gas in a 1D optical lattice}\label{sec-fermi}


Quantum degenerate fermionic atoms trapped in optical lattices \cite{mifermi} may become superfluid if there are attractive interactions between atoms trapped in two different hyperfine states \cite{fermisflat}.
Attractive fermions form pairs analogous to Cooper pairs in superconductors. A one component system of fermions trapped in the same hyperfine state may also become superfluid though not in s-wave configurations.
Such a system, in the 1D case, can be described by the Hamiltonian  ($\hbar=1$)
\begin{eqnarray}
\hat H &=& -J\sum_{j=1}^{N}(\hat c_j^\dag \hat c_{j+1}+\gamma \hat
c_j^\dag \hat c_{j+1}^\dag+h.c.-2g \hat c_j^\dag \hat c_j+g). \nonumber \\
\label{eq-ham1}
\end{eqnarray}
Here, $\hat c_j^\dag$ denotes the creation of a fermion on site
$j$,  $N$ is the number of sites, $J$ is the energy associated to
fermion tunneling to nearest-neighbor lattice sites,  $g$ is  proportional to the chemical potential
of the system and $\gamma$ accounts for the formation of pairs
within consecutive sites.  A Fourier transform shows that this
corresponds to the formation and destruction of pairs of opposite momentum  (see
\cite{xymodel,Sachdev}). A Bogoliubov transformation diagonalizes
the Hamiltonian in Eq. (\ref{eq-ham1}), which can be written up to
a zero energy shift in terms of the quasiparticle excitations $\hat d_k$,
\begin{equation}
\hat H=\sum_{k=1}^{N/2}\hat H_k= \sum_{k=1}^{N/2}E_k \hat n^d_k, \label{eq-H-d}
\end{equation}
where
\begin{eqnarray}
&\hat n^d_k= \hat d_k^\dag \hat d_k+\hat d_{-k}^\dag
\hat d_{-k} \label{eq:nk} \\
&\hat d_k=u_k \hat c_k-iv_k \hat c_{-k}^\dag, \,\,\,\,\,\,
d_k^\dag=u_k\hat{c}_k^\dag+iv_k \hat c_{-k}, \label{eq-bog} \\
& \hat c_k^\dag = \frac{1}{\sqrt{N}}\sum_{j=1}^{N}\exp(ij\Phi_k)\hat c_j^\dag \label{eq-ft} ,\\
&u_k=\cos{\frac{\theta_k}{2}}, \,\,\,\,\,\,
v_k=\sin{\frac{\theta_k}{2}}, \label{eq-ukvk}\\
&E_k=J \sqrt{(\cos{\Phi_k}-g)^2+\gamma^2\sin^2{\Phi_k}},\label{eq-epsilon} \\
& \tan{\theta_k}=\frac{\gamma\sin{\Phi_k}}{\cos{\Phi_k}-g},
\label{tantheta}
\end{eqnarray}
 and $\Phi_k=2\pi k/N$.
In order to recover the Hamiltonian (\ref{eq-ham1}), for
$(\cos\Phi_k-g) < 0$ the solution of Eq. (\ref{tantheta}) is taken
from the ($\frac{\pi}{2},\frac{3\pi}{2}$)-branch of the tangent,
whereas for $(\cos\Phi_k-g) < 0$ it is taken from the
($-\frac{\pi}{2},\frac{\pi}{2}$)-branch.

In the non-interacting case, one can clearly see that there are
two different regimes.  For $\gamma=0$ the momentum space
representation of Eq. (\ref{eq-ham1}), $\hat{H}_k=2
[\cos(\Phi_k)-g]\hat{c}_k^\dag \hat{c}_k$, is recovered up to a
constant term.  For small transverse field $g\ll1$, the energy gap
of the particles involved is positive. For high transverse field
$g \gg1$, it is negative and it vanishes at the critical point
$g=1$. It can be seen from Eq. (\ref{eq-ukvk}) that the
coefficients $u_k^2$ and $v_k^2$ change their roles at the phase
transition such that on one side of the critical point, the number
operator of the quasiparticles $\hat{d}_k^\dag \hat{d}_k$
corresponds to $\hat{c}_k^\dag \hat{c}_k$, whereas on the other
side it corresponds to $\hat{c}_k \hat{c}_k^\dag$. Finite interactions $\gamma$ between the fermions lead to the formation of fermionic pairs within consecutive sites but the main character of the phase transition at $g=1$ remains essentially unchanged .

Quantum phase transitions are only well defined at zero temperature. Thermal fluctuations lead to an exponential decay of the order parameter and only a crossover between phases remain. For the system under consideration the critical point $g=1$ at $T=0$ extends for finite $T$ to a quantum critical crossover region where the energy gap is smaller than the thermal fluctuations, i.e. $|J(1-g)|< k_BT $ \cite{Sachdev}.

The system considered here (Eq.
(\ref{eq-ham1})) is also interesting because it is equivalent to the anisotropic quantum XY
spin model \cite{xymodel}. Using the Jordan-Wigner transformation
\cite{jw,Sachdev}, one can transform it into
 \begin{equation}
H_{xy}=-J\sum_{j=1}^{N}\left[(1+\gamma)S_j^xS_{j+1}^x+(1-\gamma)S_j^yS_{j+1}^y+
g S_j^z\right], \label{eq-xy}
\end{equation}
where $S_j^\alpha$ are the spin $1/2$ operators at site $j$, $J$
is the coupling strength, $0<\gamma<1$ is the anisotropy
parameter, and $g$ is the parameter of the transverse field. The
case $\gamma=1$ corresponds to the Ising model in a transverse
field. For $\gamma=0$, the system corresponds to the isotropic
XY-model or XX-model.  For this value, the Jordan-Wigner transformation is ill-defined and one cannot map it to the fermionic Hamiltonian Eq.(\ref{eq-ham1}).  We study the phase transition with respect
to the parameter $g$, where the extreme cases $g=0$ and
$g=\infty$, correspond to  systems with no external field and with
no interactions, respectively. The phase transition between the
states with different orientations of the magnetization takes
place at $g=1$. For small transverse fields $g<1$, the ground
state has magnetic long-range order and the excitations correspond
to kinks in domain walls. For high transverse fields $g>1$, the
system is in a quantum paramagnetic state.

\section{Fermion Counting Statistics}
\label{sec_counting} Before presenting our calculations of the
counting distribution for a system of fermions at finite temperature,
we would like to remind
ourselves
of
some
basics of photon and
atom counting statistics. The theoretical analysis of the counting
process
 of photons registered on a photodetector was first developed in \cite{Glauber}.
In such a process, a photon is annihilated and a photoelectron is
emitted. This photoemission triggers a further ionization process,
leading to a macroscopic current that is then measured. This
theoretical framework can be extended for counting atoms directly
using multichannel plates or \emph{in-situ} counting techniques,
both for bosons and fermions \cite{Cahill_Glauber}. In the
detection process, the particles are absorbed by the detector. The
counting distribution can thus be derived from the master equation
that describes the interaction between the system and the detector
with efficiency $\varepsilon$ (see the Appendix). The probability
$p(m)$ of counting $m$ particles is  given by
\begin{equation}\label{eq-p}
p(m)=\frac{(-1)^m}{m!}\frac{d^m}{d\lambda^m}\mathcal{Q}\Big|_{\lambda=1},
\end{equation}
where we have used the generating function
\begin{equation}
\mathcal{Q}(\lambda)=\mbox{Tr}(\rho:e^{-\lambda\mathcal{I}}:).\label{eq-Q}
\end{equation}
Assuming that the counting process is much faster than the dynamics of the system, the time independent intensity registered at the detector is
$\mathcal{I}=\kappa\sum_{j=1}^{N} \hat c_j^\dag
\hat c_j$, where
$\kappa=1-\exp({-\varepsilon \tau})$ and $\tau$ denotes the detection exposure time.
Using the anticommutation relations for fermions we obtain
\begin{eqnarray}
 \mathcal{Q}(\lambda)
&=&\mbox{Tr}(\rho
\prod_{k=1}^{N/2}(1-\lambda\kappa \hat c_k^\dag \hat c_k)(1-\lambda\kappa
\hat c_{-k}^\dag \hat c_{-k})).\label{eq-Q-a}
\end{eqnarray}
The dynamics mixes only $k$ and $-k$ fermionic excitations $\hat d_k$ so that we can separate the density matrix $\rho=\prod_k \rho_k$ and neglect the terms which do not conserve the number of excitations to obtain
\begin{eqnarray}
&&\mathcal{Q}(\lambda)=\prod_{k=1}^{N/2}\big(1-\lambda\kappa
A_k+\lambda^2\kappa^2B_k\big),\label{eq-Q-d}
\end{eqnarray}
where
\begin{eqnarray}
A_k&=& \hbox{Tr}(\rho_k \left [u_k^2 \hat n^d_k+v_k^2(\hat d_k \hat d_k^\dag +\hat d_{-k}\hat d_{-k}^\dag) \right] )\nonumber \\
B_k&=&\hbox{Tr}(\rho_k \left[ u_k^2\hat d_k^\dag \hat d_k \hat d_{-k}^\dag \hat d_{-k}+v_k^2
\hat d_{-k} \hat d_{-k}^\dag \hat d_k \hat d_k^\dag \right]).\label{eq-Bk}
\end{eqnarray}
We use Eq. (\ref{eq-p}) to calculate the counting distribution
from the generating function in Eq. (\ref{eq-Q-d}) and obtain
\begin{equation}
p(m)=\frac{(-1)^m}{m!}\frac{d^m}{d\lambda^m}\Big[\prod_{k=1}^{N/2}\big(1-\lambda\kappa
A_k+\lambda^2\kappa^2B_k\big)\Big]_{\lambda=1}.\label{eq-p-A-B}
\end{equation}
Using the generalized Leibniz rule,  we derive \cite{counting} a recurrence relation to calculate the counting
distribution for a system with $M+1$ pairs of modes from the
distribution of a system with $M$ pairs of modes
\begin{eqnarray}
p(m,M+1)&=&  \sum_{i=0}^{2} {\cal P}_ip(m-i,M). \label{asol_jinis}
\end{eqnarray}
Here ${\cal P}_i$ denotes the probability of detecting $i$
particles in the two modes $M+1$ and $-(M+1)$ that is given by
\begin{eqnarray}
{\cal P}_0 &= & 1-\kappa A_{M+1}+\kappa^2B_{M+1}, \nonumber \\
{\cal
P}_1 &=& \kappa A_{M+1}-2\kappa^2B_{M+1} , \nonumber \\
{\cal P}_2 &=& 1-{\cal P}_0-{\cal P}_1.\label{eq-Ps}
\end{eqnarray}
Using the recursive relation Eq. (\ref{asol_jinis}), the counting
distribution for an arbitrarily large system can be calculated
from the counting distributions of a two mode system.  We thus
only need to calculate the expressions $A_k$ and $B_k$
 in Eq. (\ref{eq-Bk}) and use Eqs. (\ref{asol_jinis}-\ref{eq-Ps}) to obtain the counting distributions of the fermionic system Eq.(\ref{eq-ham1}) with an arbitrary number of sites.

As mentioned above, the fermionic
operators are related to spin operators by the Jordan-Wigner
transform. The fermion counting distribution is therefore, up to a
constant, equivalent to the counting distribution of the spins in
$z$-direction in the transverse XY-model in Eq. (\ref{eq-xy}).  We can thus use the above to calculate the counting distributions of the anisotropic XY model in a transverse field for a system of any size $N$.
Experimentally, the spin number distribution and its fluctuations can be inferred from the expectation value and fluctuations of the polarization of the light that has interacted with a cold atomic sample \cite{Sorensen}. This spin polarization spectroscopic technique can be also extended to account for spatial resolution of the spin distributions \cite{tobenat}.

\section{Counting statistics in the presence of thermal noise} \label{sec-counting_equi}
In real counting experiments, there are typically a variety of
noise sources that may affect the system. In this section we study
the influence of thermal noise on the counting distributions of
the 1D fermi system in Eq. (\ref{eq-ham1}). We analyze the
counting distributions along the crossover between the different
regions of the phase diagram. We first review the results for the
zero temperature case and then turn our discussion to the case
with thermal fluctuations.

\subsection{Counting statistics at zero temperature} \label{sec-counting_zero}
\begin{figure}
\centering \epsfig{file= 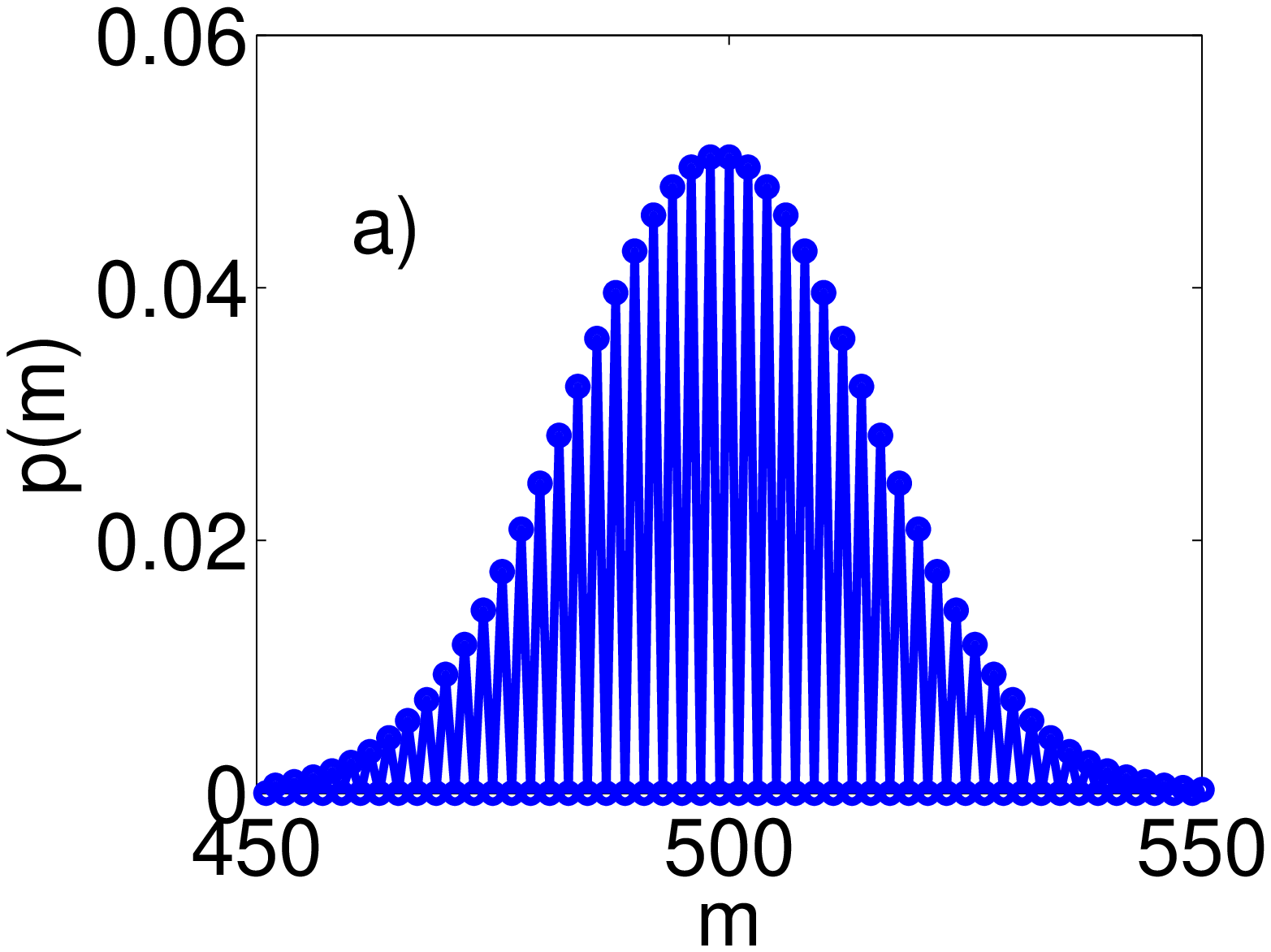,width=0.48\linewidth}
\epsfig{file= 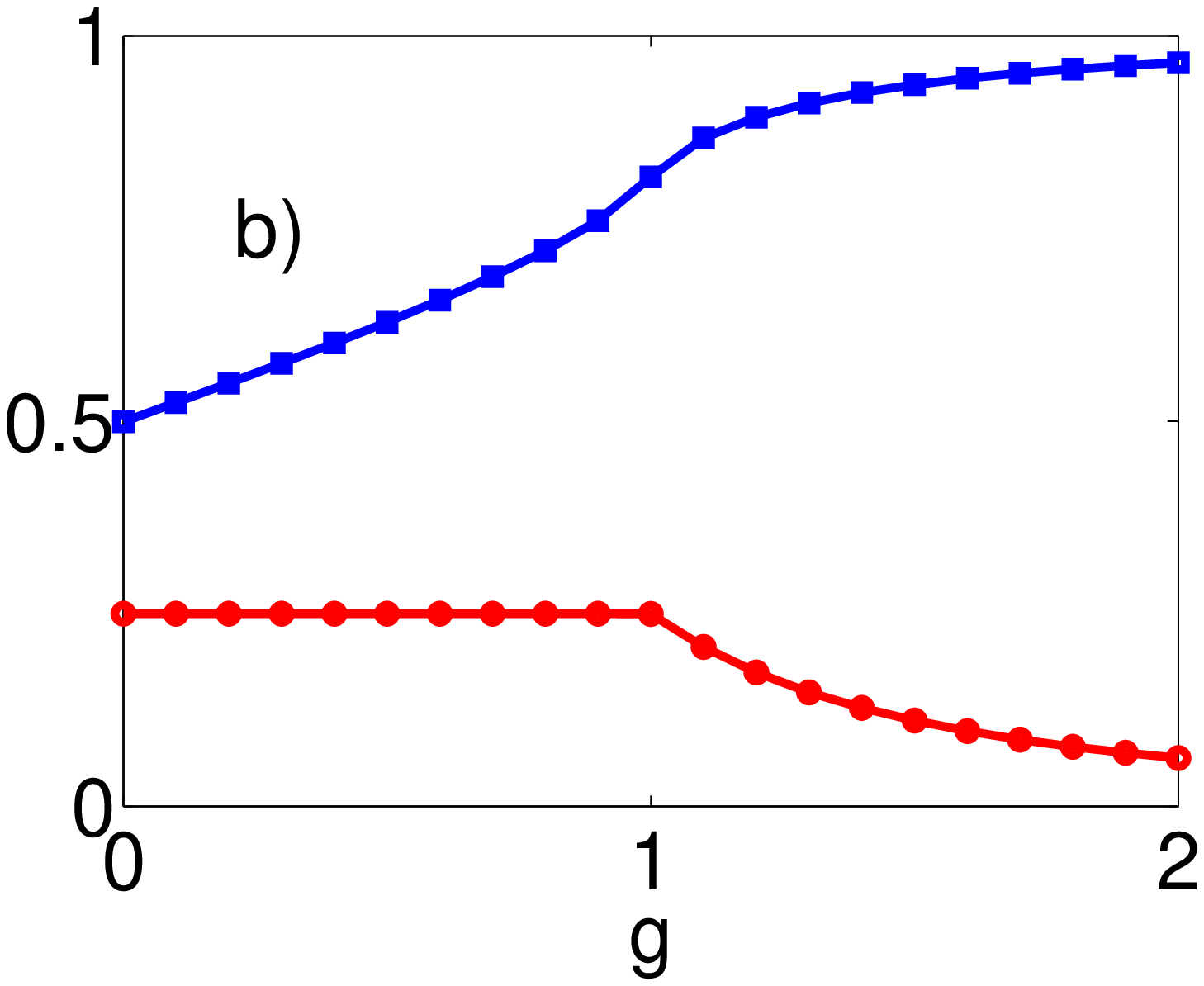,width=0.48\linewidth} \caption{a)
Counting probability distribution $p(m)$ of finding $m$ particles
as a function of $m$ for the fermionic system Eq.(\ref{eq-ham1})
with $\gamma=\kappa$=1 and $g=0$, $N=1000$ at $T=0$. b) Mean $\bar{m}/N$
(blue squares) and variance $\sigma^2/N$ (red circles) of the
counting distribution  as a function of the transverse field $g$
at T=0.
}
 \label{fig-T0}
\end{figure}

At zero temperature, the ground state of the system Hamiltonian
Eq. (\ref{eq-H-d}) is the vacuum state of $\hat d_k$ excitations.
The expressions $A_k$ and $B_k$ in Eq. (\ref{eq-Bk}) are thus
given by
\begin{eqnarray}
A_k=2\kappa v_k^2\nonumber \\
B_k=\kappa^2 v_k^2.\label{eq-Ak-Bk-vac}
\end{eqnarray}
Inserting these into the equations for the two-mode probabilities
${\cal P}_i$ in Eq. (\ref{eq-Ps}), we obtain the probabilities of
finding $0$, $1$ or $2$ particles in a system with one pair of
modes
\begin{eqnarray}
p(0,1)=1-2\kappa (v_{1}^2+\kappa^2v_{1}^2)\nonumber\\
p(1,1)=2\kappa v_{1}^2-2\kappa^2v_{1}^2\nonumber\\
p(2,1)=\kappa^2v_{1}^2.
\end{eqnarray}
The counting distribution can now be calculated for an arbitrary
number of modes using the recurrence relation Eq.
(\ref{asol_jinis}). In fig. \ref{fig-T0} a , we plot the counting
probability distribution for the Ising model ($\gamma=1$) at zero
temperature with no transverse field ($g=0$) and perfect detection
efficiency. We consider a system with zero excitations and
$N=1000$ sites. The probability distribution is centered around a
mean value $\bar{m}=500=N/2$ particles and its standard deviation
$\sigma=50$ such that $\sigma^2=N/4$. As was observed in \cite{counting}, the pairing that
is present in the system hamiltonian Eq. (\ref{eq-ham1}) only
allows for the detection of pairs of particles and thus leads to a
zero probability of finding an odd number of particles. In
\cite{counting}, this splitting of the counting distribution
between even and odd values was shown to disappear for decreasing
detector efficiency $\kappa$. In the next section Sec. \ref{sec-fert}, we use this
feature of the counting distribution to study the influence of thermal fluctuations
on the stability of the fermion pairs. In fig.
\ref{fig-T0} b, we plot the mean $\bar{m}$ and variance
$\sigma^2$ of the counting distribution
 for different values of the transverse field $g$. The mean number of particles
increases with increasing transverse field $g$. The variance is
constant with $g$ up to the critical point, when it decreases with
increasing $g$. The phase transition at $g=1$ is clearly visible
both in the mean and in the variance. In \cite{counting}, we
studied the behavior of the counting distribution for different
values of the anisotropy parameter $\gamma$ and the detection
efficiency $\kappa$. We found that the characteristic behavior of
the mean and variance as shown in fig~\ref{fig-T0} b) is similar
when $\gamma$ varies from $0$ to $1$. We further found that the
phase transition is visible in the means and variances even for
small detection efficiencies. In the following we
consider full detection efficiency ($\kappa=1$), as the results
for smaller efficiencies are similar.

\subsection{Counting statistics at non-zero temperature} \label{sec-fert}
We now turn our discussion to the case of non-zero temperature.  The effect of thermal fluctuations in the system we consider is two-folded. On the one hand, thermal fluctuations allow for breaking of superfluid fermionic pairs. On the other hand, the quantum phase transition reduces to a crossover between different regions of the phase diagram. We will show that both effects are visible in the counting distribution functions.

We consider the counting statistics at finite temperature $T$
using the canonical ensemble,
$\rho=\frac{1}{Z}e^{-\beta \hat H}$,
where $\beta=\frac{1}{k_B T}$, the Hamiltonian is given by Eq. (\ref{eq-H-d}) and the
partition function $Z=\mbox{Tr}\left(e^{-\beta \sum_k \hat H_k}\right)$. The finite temperature $T$ determines the average number of quasiparticle excitations $\hat d_k$.
 In order to calculate the terms $A_k$ and $B_k$ defined in Eq.
 (\ref{eq-Bk}), we write
$\rho_k=\frac{1}{Z_k} e^{-\beta \hat H_k}$ where $Z_k=\mbox{Tr}\left(e^{-\beta \hat H_k}\right)$ and we take the
trace in the basis $\{|00\rangle |01\rangle |10\rangle
|11\rangle\}$.
We obtain
\begin{eqnarray}
A_k=\frac{2\kappa}{Z_k}
(v_{k}^2+{e^{-\beta E_{k}}}
+e^{-2 \beta E_{k}}u_{k}^2)\nonumber\\
B_k=\frac{\kappa^2}{Z_k}(v_{k}^2+e^{-2 \beta E_{k}}
u_{k}^2)\label{eq-Ak-Bk-T} \\
Z_k=1+2{e^{-\beta E_k}}+{e^{-2 \beta E_k}}
\end{eqnarray}
For a given value of the transverse field $g$, we fix the
temperature and obtain the number $N_d=\sum_{k=1}^{N/2}N^d_k $ of
fermionic excitations
\begin{equation}
 N^d_k=\mbox{Tr}\left(\rho_k \hat n^d_k \right). \label{eq-Nk}
\end{equation}
As explained above, we use $A_k$ and $B_k$ to obtain the recursive
formula for the counting distribution. \\

\begin{figure}
\centering
\epsfig{file= 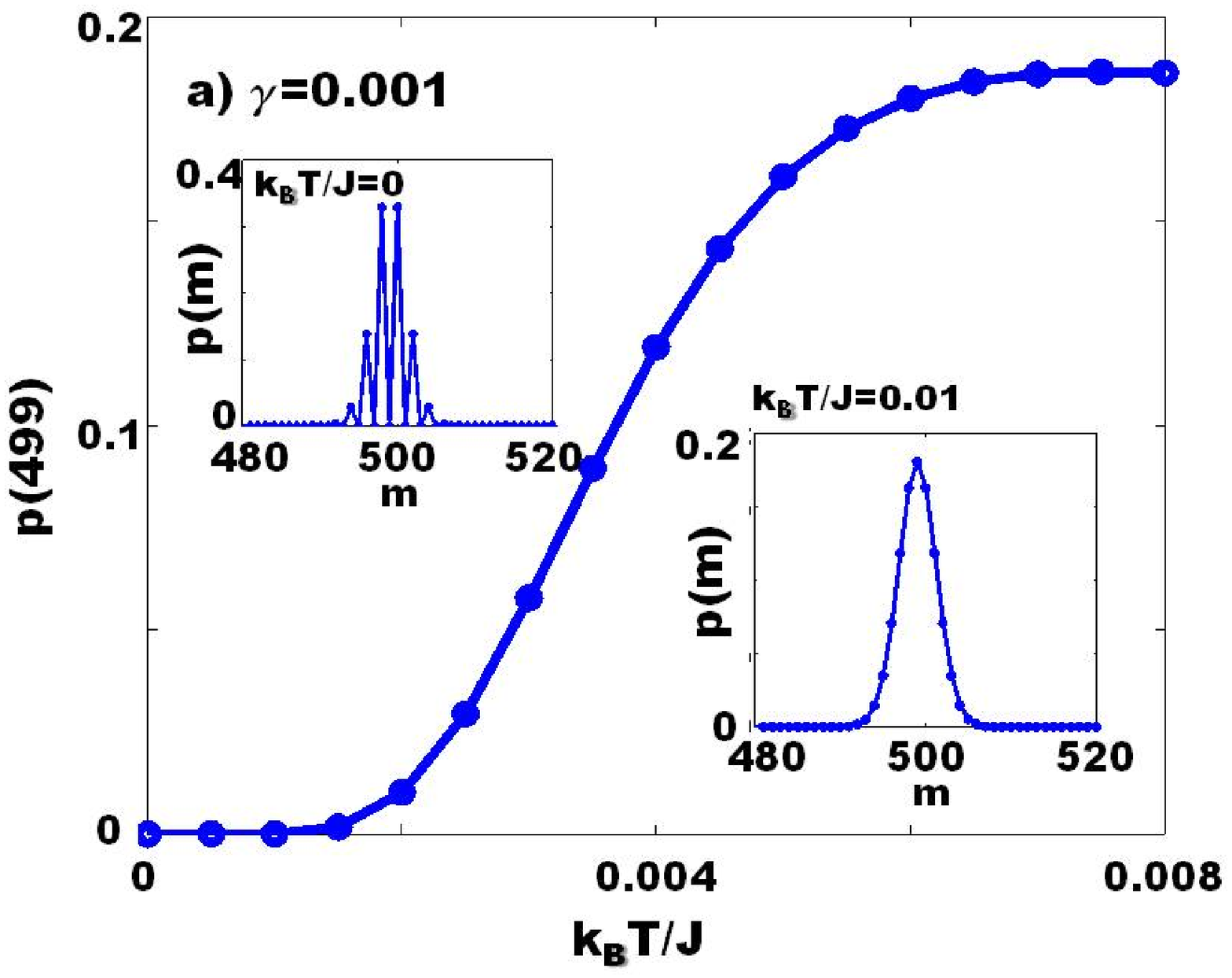,width=\linewidth} \epsfig{file=
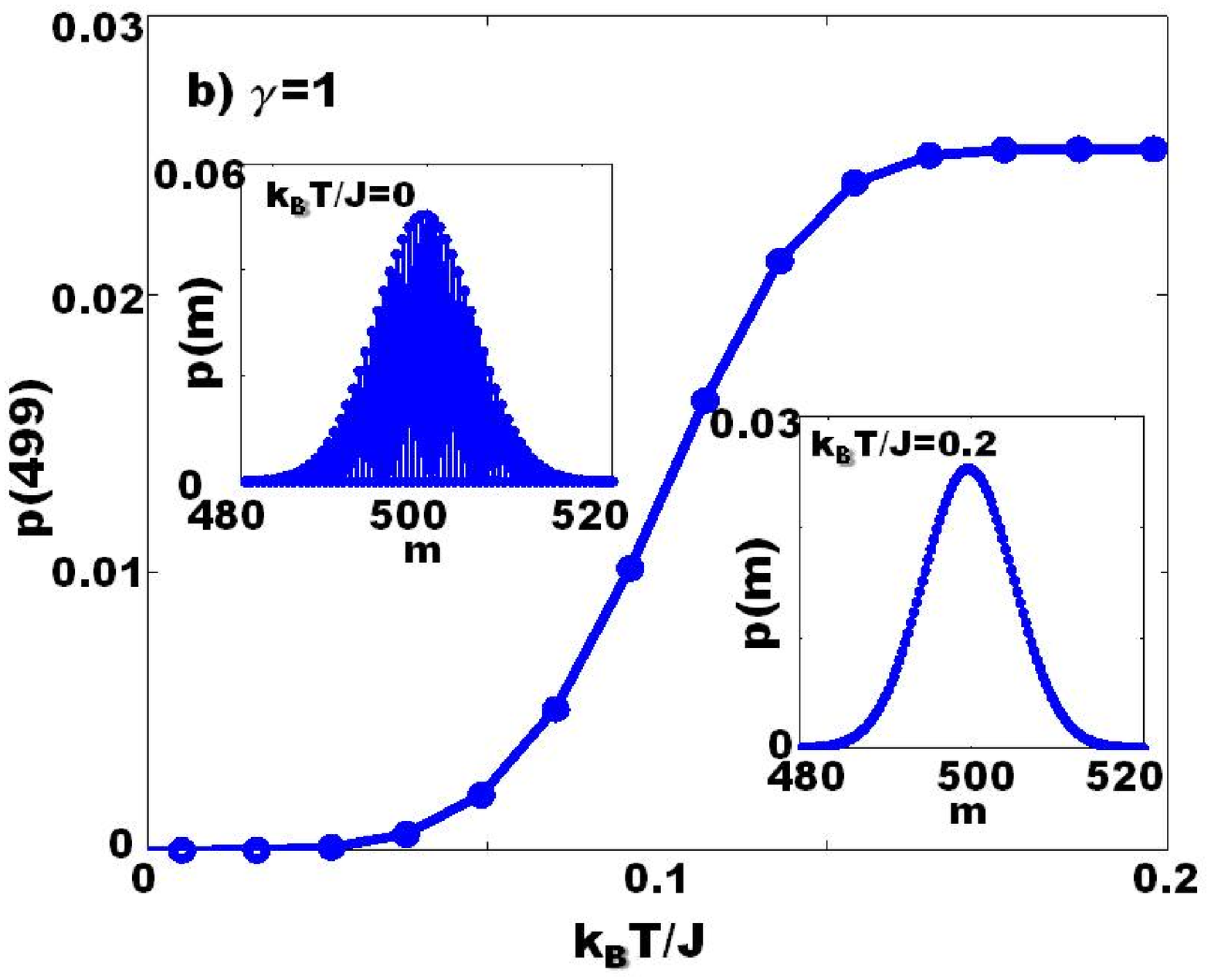,width=\linewidth} \caption{Probability of counting
an odd number (m=499) of particles as a function of T for
$\gamma=0.01$ (fig. a) and $\gamma=1$ (fig. b). The insets show
the counting distribution for $T=0$ and $k_BT/J=0.01$ in fig. a) and for
$T=0$ and $k_BT/J=0.2$ in fig. b).} \label{fig-pairs}
\end{figure}

Thermal fluctuations induce the breaking of pairs. For increasing temperature, the pairing of fermions whose binding energy is proportional to $\gamma$ in Eq.
(\ref{eq-ham1}) is suppressed. This is reflected in the counting distribution in such a way that the counting probability
for odd numbers of particles becomes non-zero. To illustrate this,
 we plot in fig. 2 the probability of counting the
exemplary odd value of $m=499$ particles as a function of
temperature. As temperature increases, the pairs are destroyed and
we observe a transition from zero probability to a finite value.
We compare a system with small interaction strength $\gamma=0.01$
(fig. \ref{fig-pairs} a ) to the case of $\gamma=1$ (fig.
\ref{fig-pairs} b ).  In the insets, we compare the counting
distribution for each system at zero temperature and at higher
temperatures. We observe that the splitting between even and odd
particle numbers disappears as the temperature increases. Note
that here we consider a perfect detection process. For lower
detection efficiency, the splitting is not visible, as was shown
in Ref. \cite{counting}. For small interaction strength $\gamma$,
the counting distribution is narrower, while higher binding
energies $\gamma$ imply broader atom number distribution
functions. Also, observing the scales of temperatures when the
counting of odd particles become non-zero, one can
infer that this temperature is proportional to the parameter $\gamma$. \\

\begin{figure}
\centering
\begin{tabular}{ccc}
\epsfig{file= 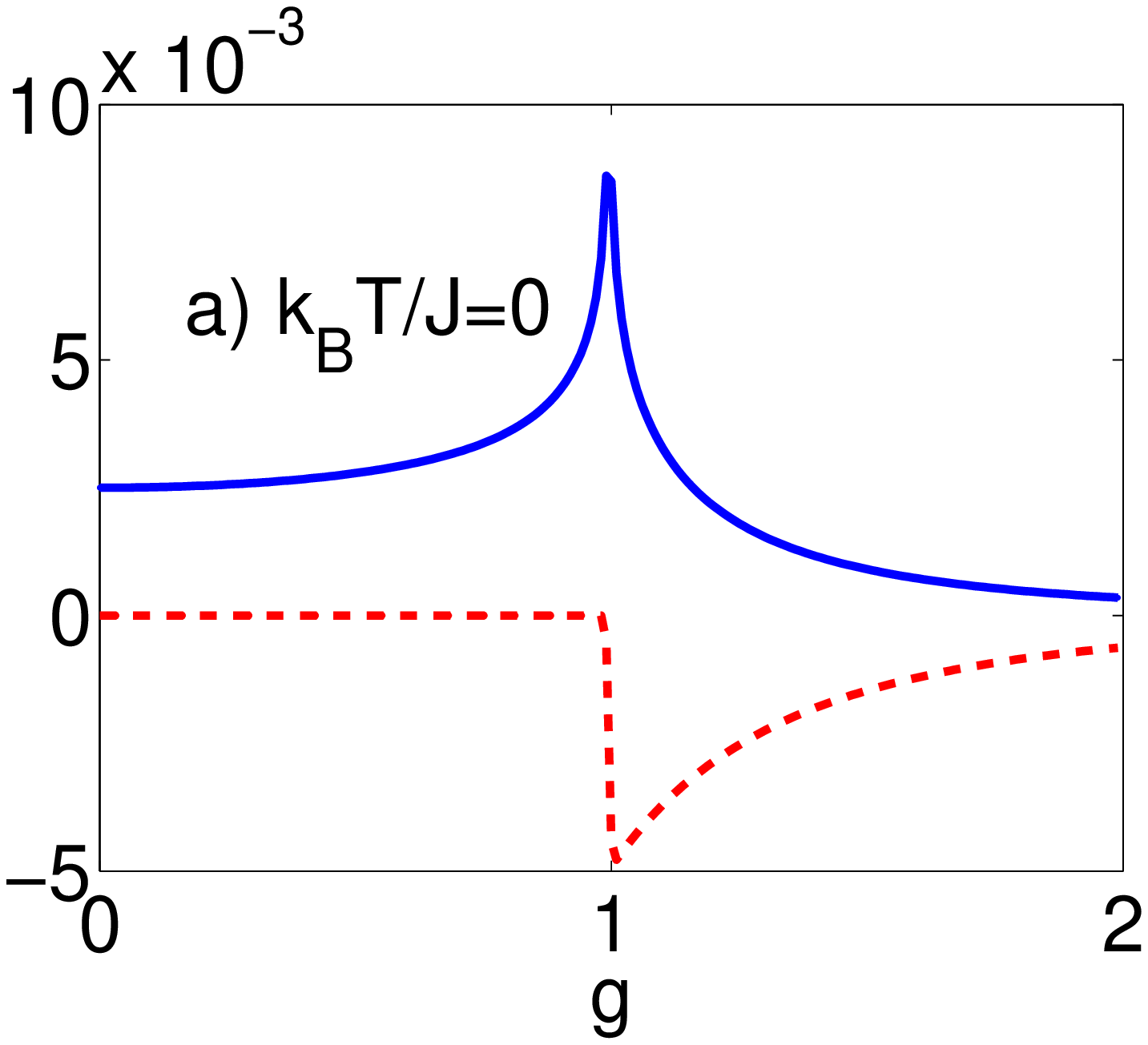,width=0.5\linewidth} &\epsfig{file=
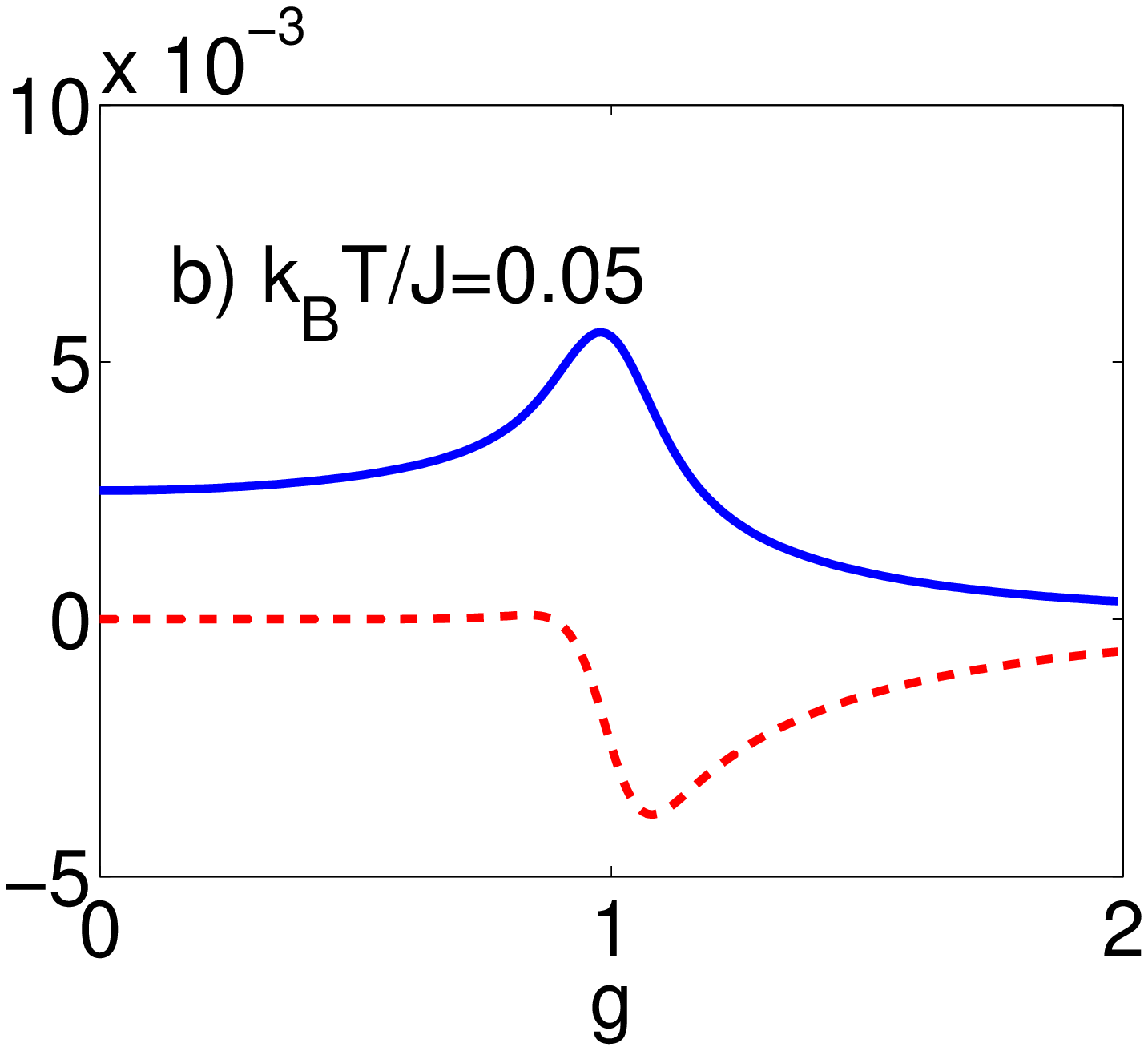,width=0.5\linewidth}
\\\epsfig{file= 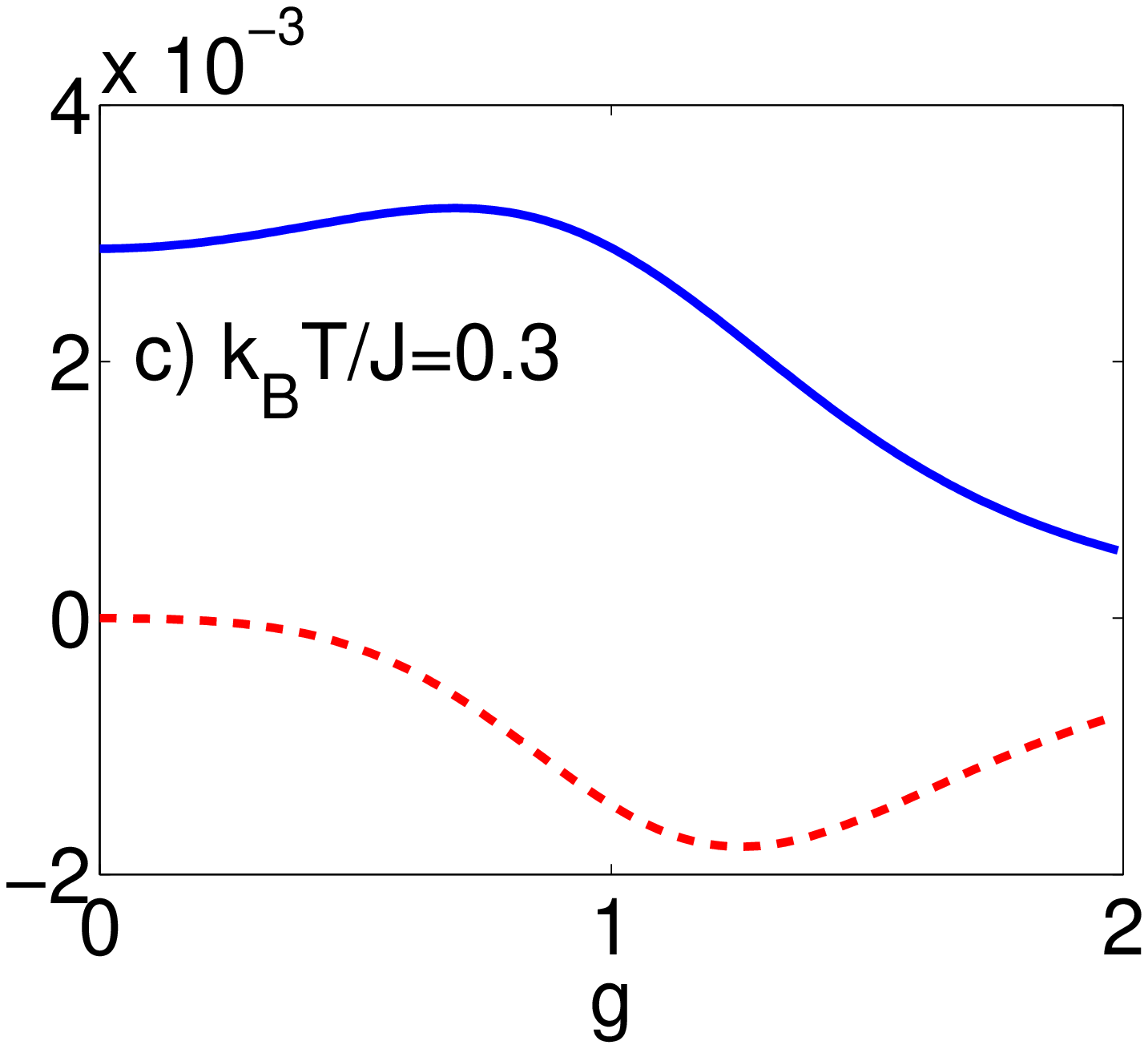,width=0.5\linewidth} & \epsfig{file=
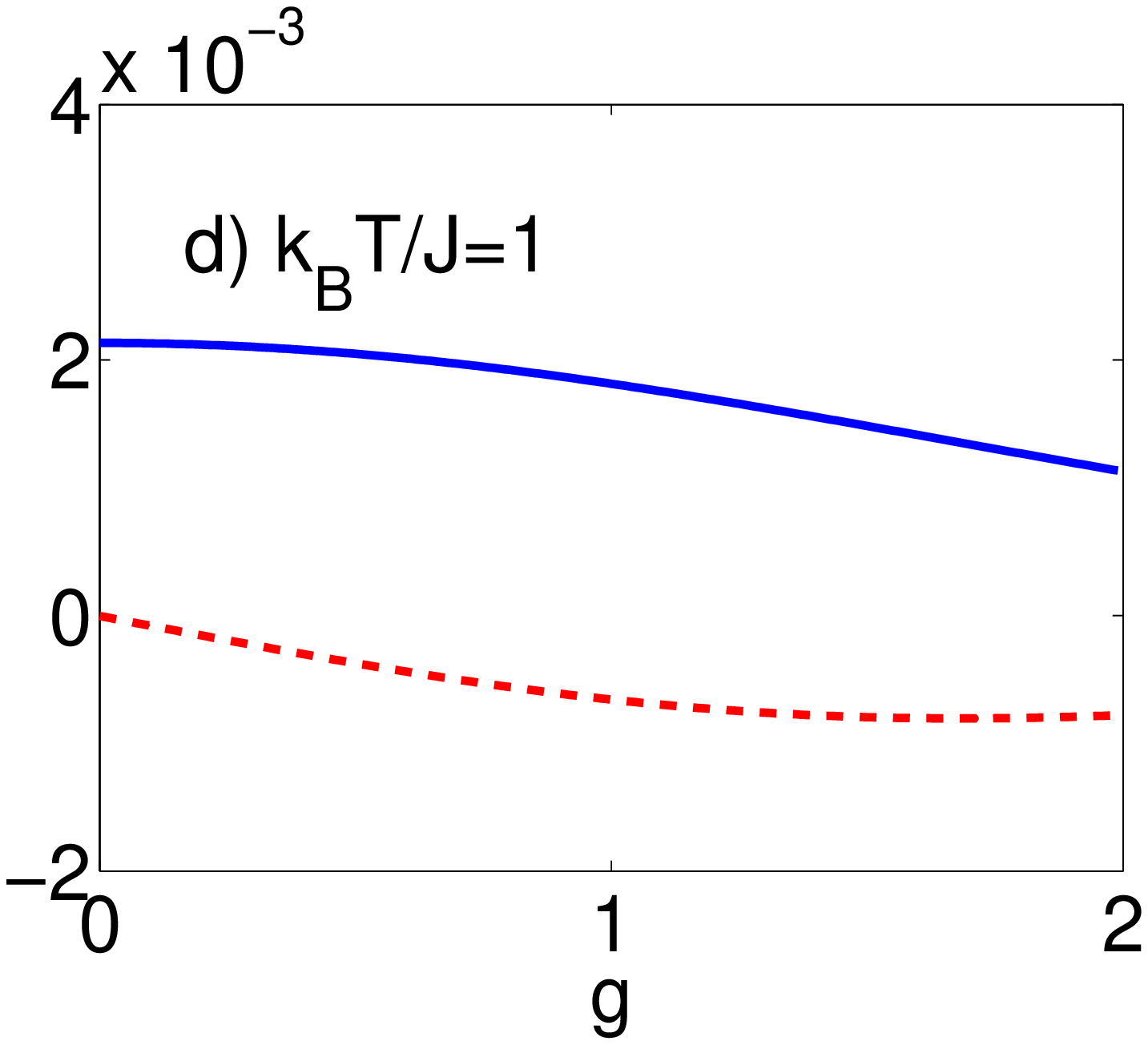,width=0.5\linewidth}
\end{tabular}
\caption{Derivative of the mean $\bar{m}/N$ (blue solid line) and
the variance $\sigma^2/N$ (red dashed line) of the counting
distribution of the fermionic system Eq.(\ref{eq-ham1}) with
$\gamma=1$ as a function of the transverse field $g$. a) $T=0$,
$N_d=0$ for all $g$; b) $k_BT/J=0.05$ and $N_d/N\simeq 0$ at
$g=0$; c) $k_BT/J=0.3$ and $N_d/N=0.03$ at $g=0$; d) $k_BT/J=1$
and $N_d/N=0.27$ at $g=0$. } \label{fig-mean-var-Temp}
\end{figure}

Let us now turn our discussion to the influence of temperature on
the criticality of the system. As was seen above for the case of
zero temperature, the phase transition is visible in the mean and
variance of the distribution. This behavior is even more evident
in the derivatives of the mean and variance. In fig.
\ref{fig-mean-var-Temp}, we plot the derivative of the means and
variances with respect to g at different temperatures $T$. One can
see how the criticality is blurred when the temperature is of the
order of the energies of the system $k_B T\sim E_k$. At high
temperature, the mean and variance become independent of the
transverse field value $g$ and take a constant value of $0.5N$ and
$0.25N$, respectively.

\section{Counting statistics during thermalization of a system coupled to a heat bath} \label{sec-counting_heat}
The long decoherence times of experiments with ultracold atoms
allow to study the real time quantum dynamics of the system. The
dynamics of an open system coupled to a heat bath has recently
aroused much interest \cite{env} as one can use dissipation for
quantum state engineering. By tuning the properties of the
reservoir, thermalization can drive the system to a steady state
which has the desired properties and can e.g. be used to encode
quantum information. Here, we consider the thermalization of the
system hamiltonian Eq. (\ref{eq-ham1}), when it is coupled to a
heat bath. We start from the ground state at $T=0$ and let the
system evolve to the thermal Boltzmann-Gibbs equilibrium state. In
this sense, we analyze the counting statistics in a temperature
quench. Coupling to the heat bath is described by the quantum master
equation \cite{breuer}
\begin{eqnarray}
&&\frac{d}{dt}\rho(t)=\nonumber\\&&\gamma_0\sum_k
(\frac{N^d_k}{2}+1) \left [ \hat d_k\rho(t)\hat
d_k^\dag-\frac{1}{2}\hat d_k^\dag \hat
d_k\rho(t)-\frac{1}{2}\rho(t) \hat d_k^\dag \hat d_k\right]
\nonumber\\&&+ \gamma_0 \sum_k  \frac{N^d_k}{2} \left[ \hat
d_k^\dag\rho(t) \hat d_k-\frac{1}{2}\hat d_k \hat d_k^\dag
\rho(t)-\frac{1}{2}\rho(t)\hat d_k \hat
d_k^\dag\right],\label{eq-master-heat}
\end{eqnarray}
where $\gamma_0$ is the coupling strength and $N^d_k$ defined in Eq.
(\ref{eq-Nk}), accounts for the mean number of fermions in the mode
$k$ at a certain temperature $T$. This open system dynamics
assures that the system approaches thermal equilibrium towards the
Boltzmann-Gibbs state.

At this point, we would like to
clarify an important point in relation to particle counting of a dynamical system. The system is governed by two
different dynamic processes, one is the coupling to the heat bath
described by Eq. (\ref{eq-master-heat}), the other one is the
detection by particle counting described by Eq. (\ref{master}) in the Appendix. We
assume that the coupling of the system to the heat bath occurs on a time scale much
slower than the counting process. The counting is thus performed
in a time interval in which the coupling to the bath does not
affect the system, so that it can be considered time independent.
Below we show how the counting statistics change
during thermalization of the system with the heat bath. However, each of the
distributions is registered at the detector in a time interval in which no change
occurs.

\subsection{Coupling to the excitations} \label{sec-non-local}In order to
calculate the counting statistics of the system coupled to a heat
bath, we calculate the terms $A_k$ and $B_k$ as given in Eq.
 (\ref{eq-Bk}), which now depend on time.
From the master equation (\ref{eq-master-heat}), the time
dependent mean excitation number is obtained as
\begin{equation}
\langle \hat n_k^d (t) \rangle=\hbox{e}^{-\gamma_0 t} \langle   \hat n_k^d (0)\rangle +N^d_k(1-\hbox{e}^{-\gamma_0 t}).
\end{equation}
We start with the system initially in the vacuum state and use
\begin{eqnarray}
\langle \hat d_k^\dag \hat d_k \hat d_{-k}^\dag \hat d_{-k}  (t) \rangle=\langle \hat d_k^\dag
\hat d_k\rangle_t\langle \hat d_{-k}^\dag \hat d_{-k}\rangle_t,
\end{eqnarray}
to calculate the time dependent terms $A_k(t)$ and $B_k(t)$ for a
system in a heat bath
\begin{eqnarray}
&\frac{A_k(t)}{\kappa}=u_k^2N^d_k(1-e^{-\gamma_0
t})+v_k^2(2-N^d_k(1-e^{-\gamma_0 t})),\nonumber
\\
&\frac{B_k(t)}{\kappa^2}=u_k^2\frac{(N_k^d(1-e^{-\gamma_0
t}))^2}{4}\nonumber\\&+v_k^2(1-N_k^d(1-e^{-\gamma_0
t})+\frac{(N_k^d(1-e^{-\gamma_0 t}))^2}{4}).
\end{eqnarray}

\begin{figure}
\centering
\epsfig{file= 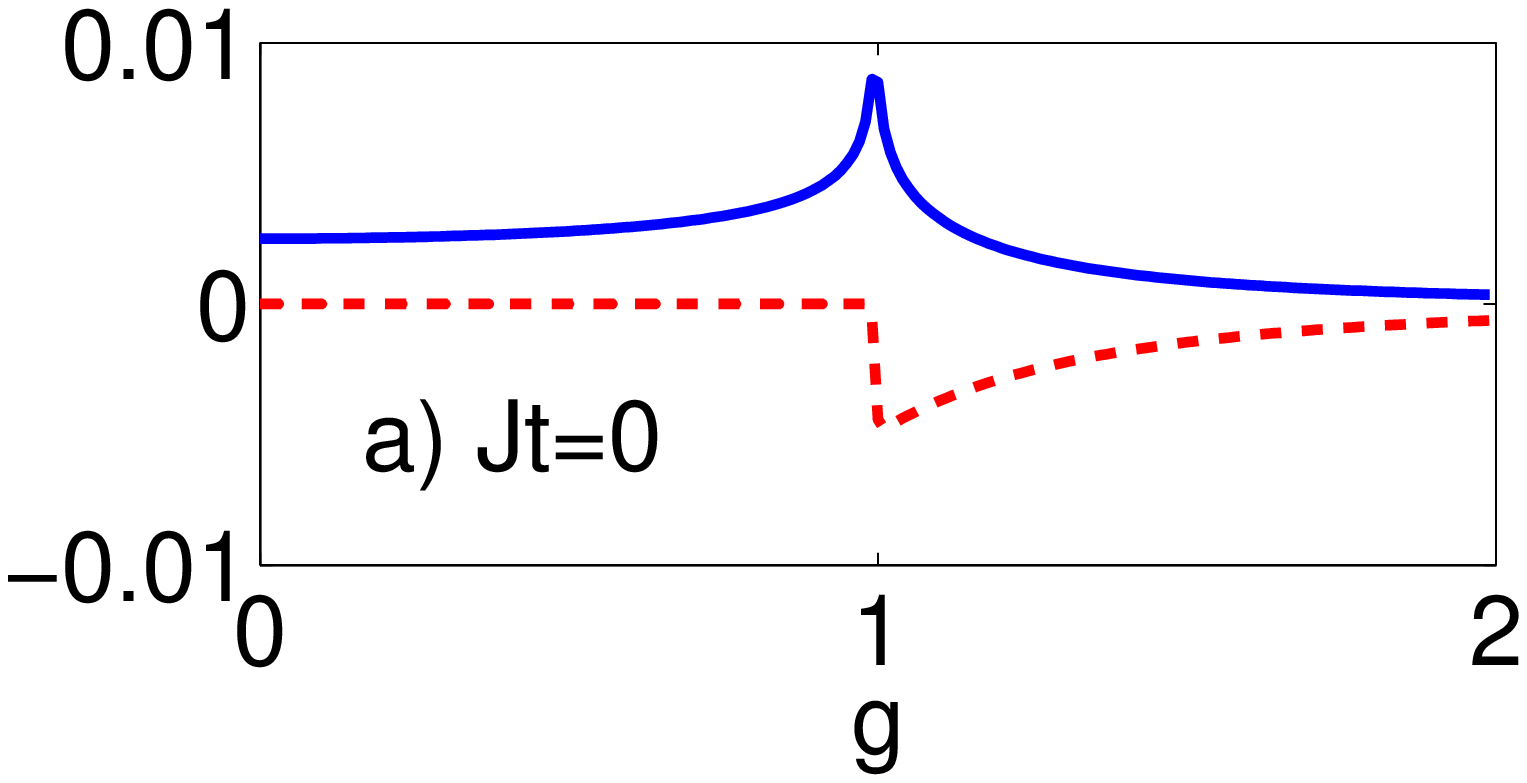,width=0.8\linewidth} \epsfig{file=
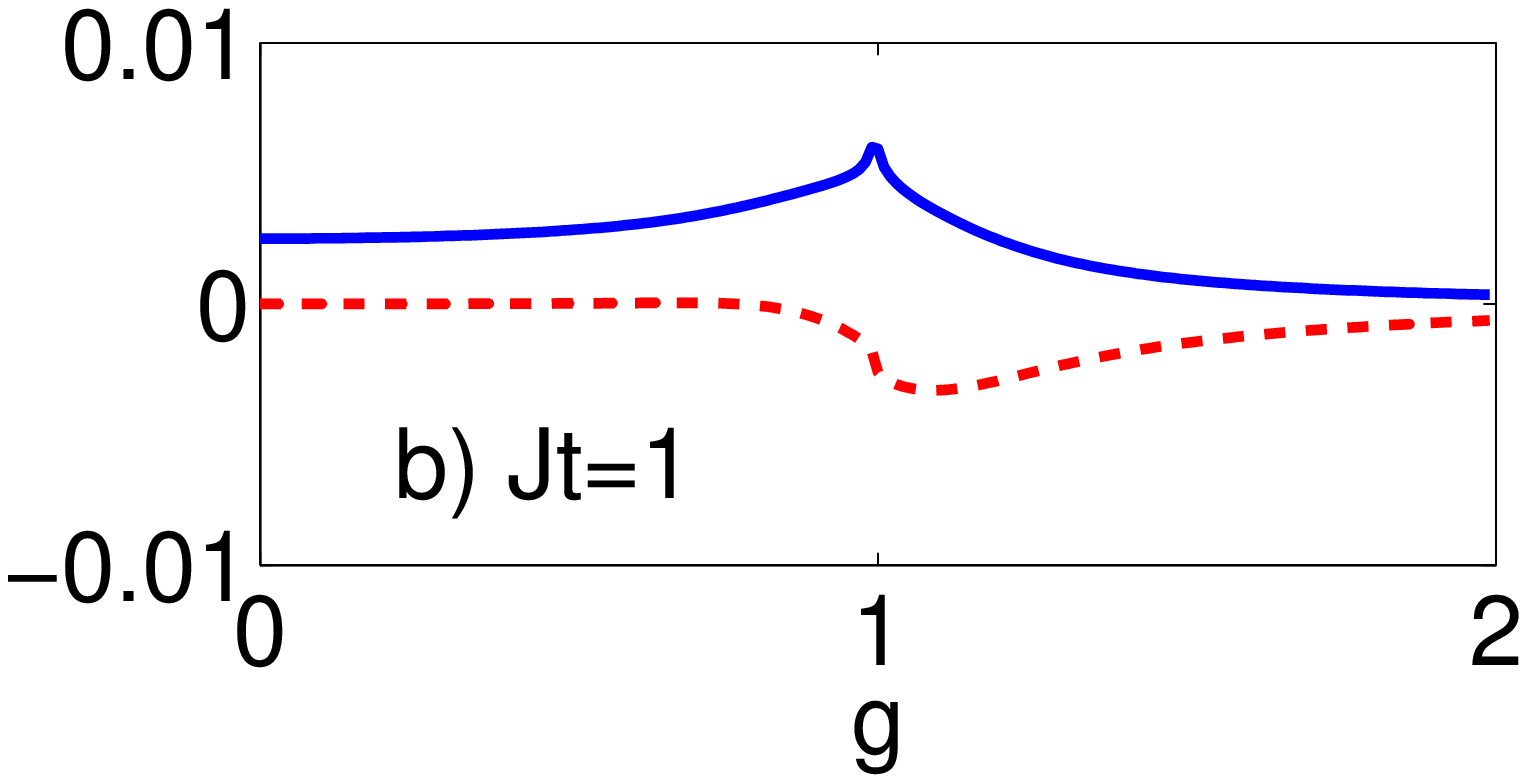,width=0.8\linewidth} \epsfig{file=
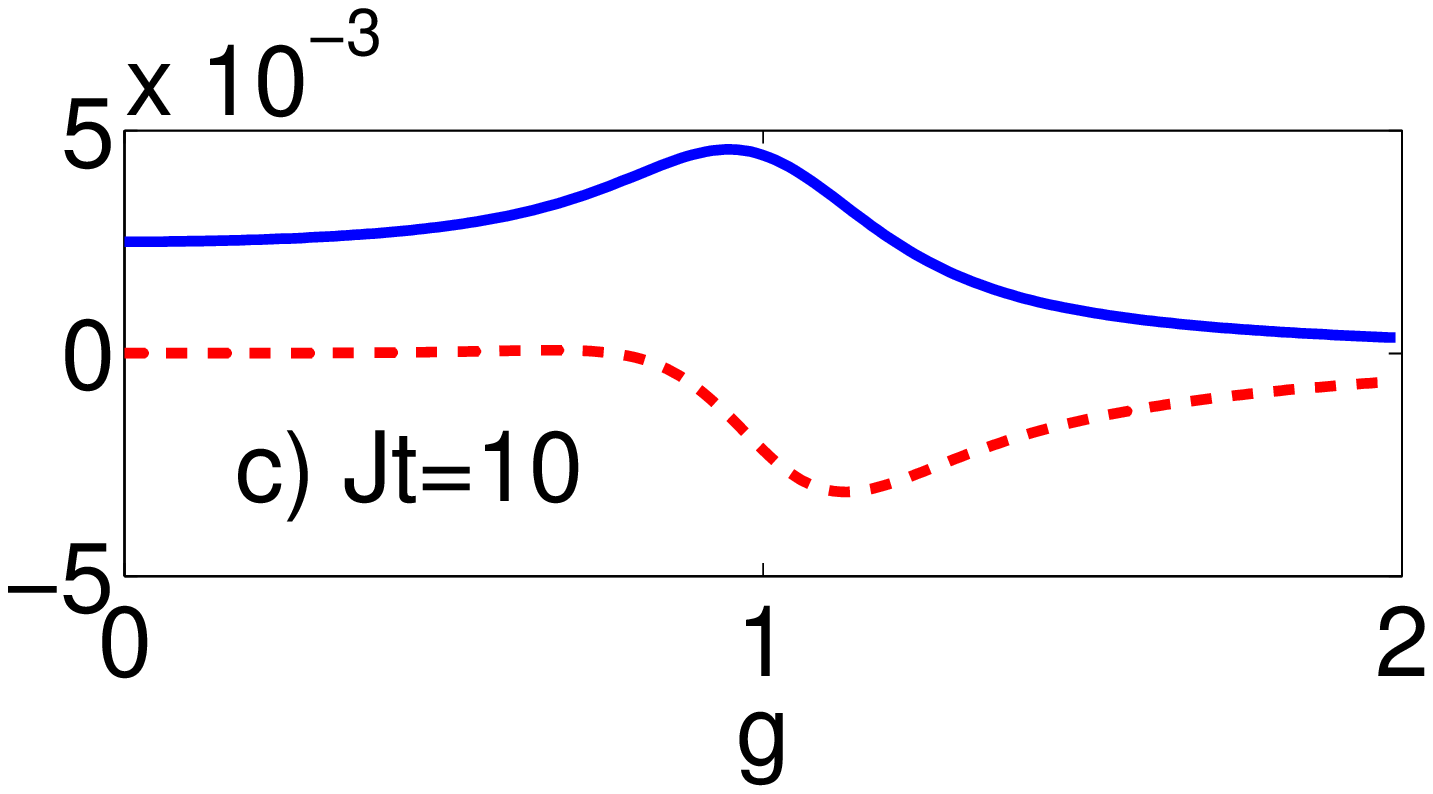,width=0.8\linewidth}
\caption{Thermalization: Derivative with respect to the parameter
$g$ of the mean (blue solid line) and variance (red dashed line)
for $\gamma=1$ for increasing coupling time with $\gamma_0$=1 and
$k_BT/J=0.1$. Fig. a) shows the initial time when the system is
not coupled to the bath. Fig. b) $J t=1$ and c)$J t=10$. }
\label{heatbath}
\end{figure}

In fig. \ref{heatbath}, we plot the derivatives of the mean and
variance with respect to the transverse field $g$ at different times $t$ at a fixed
coupling rate $\gamma_0=1$ and at fixed temperature of the bath
$k_BT/J=0.1$. At the initial time $t=0$, the mean and variance
correspond to those of the zero excitation state, ground state at
zero temperature (fig. \ref{heatbath} a ). The phase transition is
clearly visible in the derivative both of the mean and the
variance. Due to the coupling of the system and the bath, already
for intermediate times (see fig. \ref{heatbath} b ), the
characteristic behavior of the mean and variance in the critical region washes out. For long coupling time, as shown in fig.
\ref{heatbath} c , the behavior is completely determined by the
bath.

\begin{figure}
\centering
\epsfig{file= 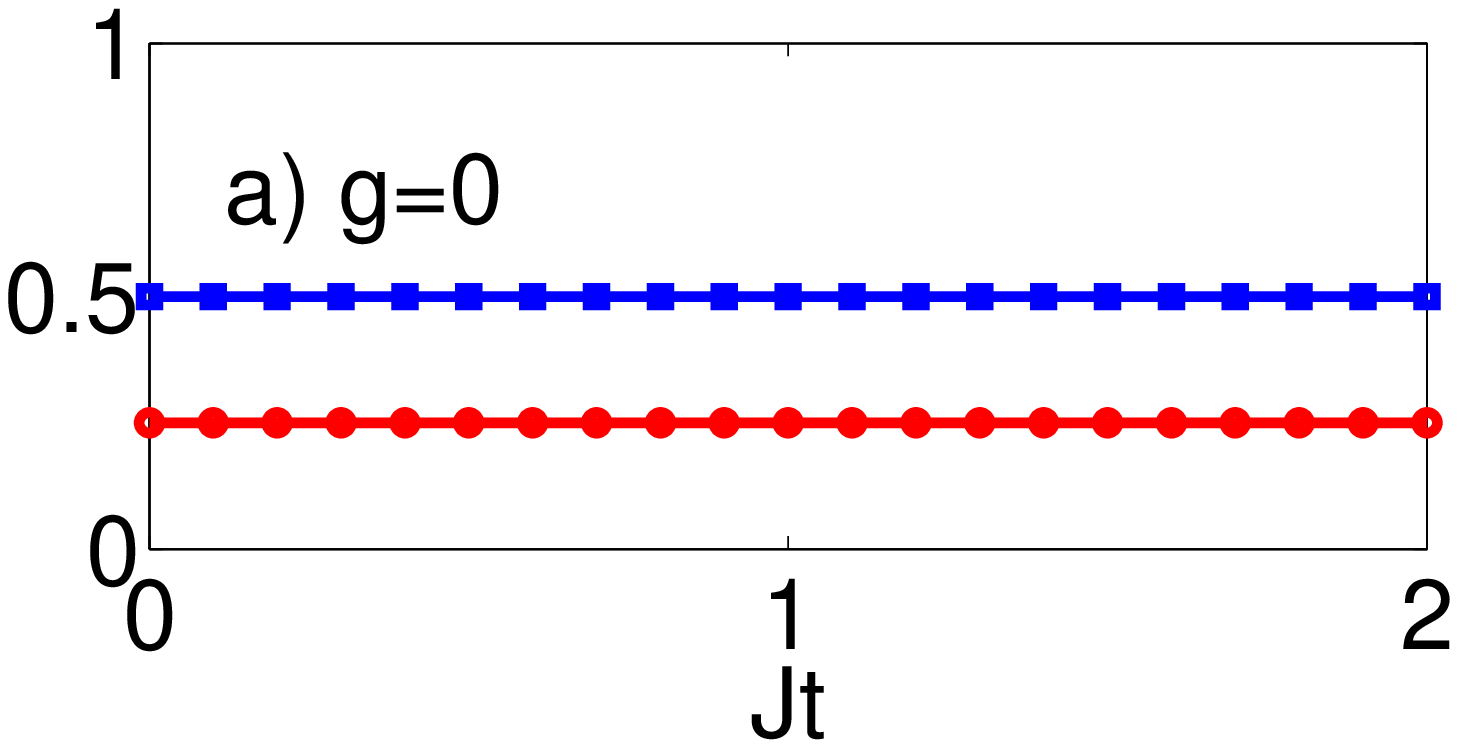,width=0.8\linewidth} \epsfig{file=
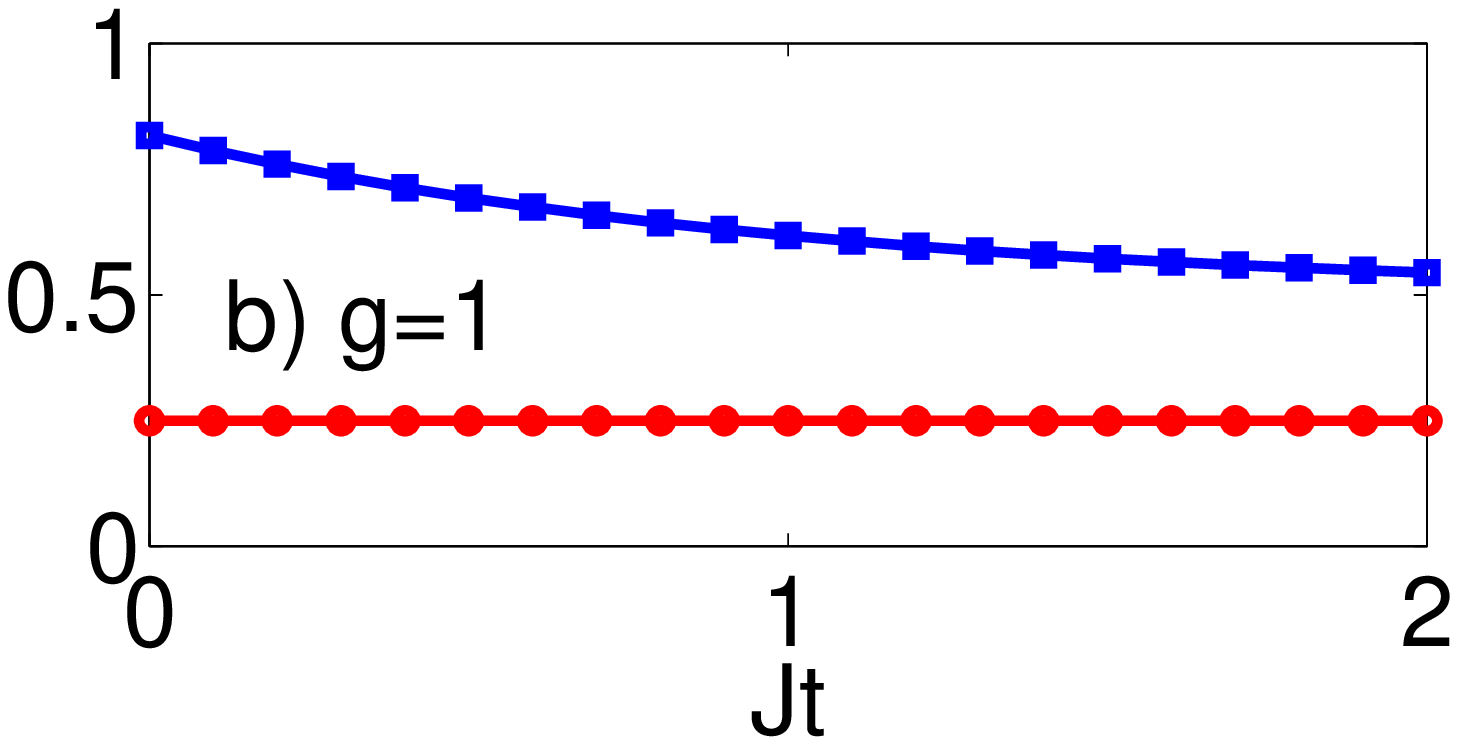,width=0.8\linewidth} \epsfig{file=
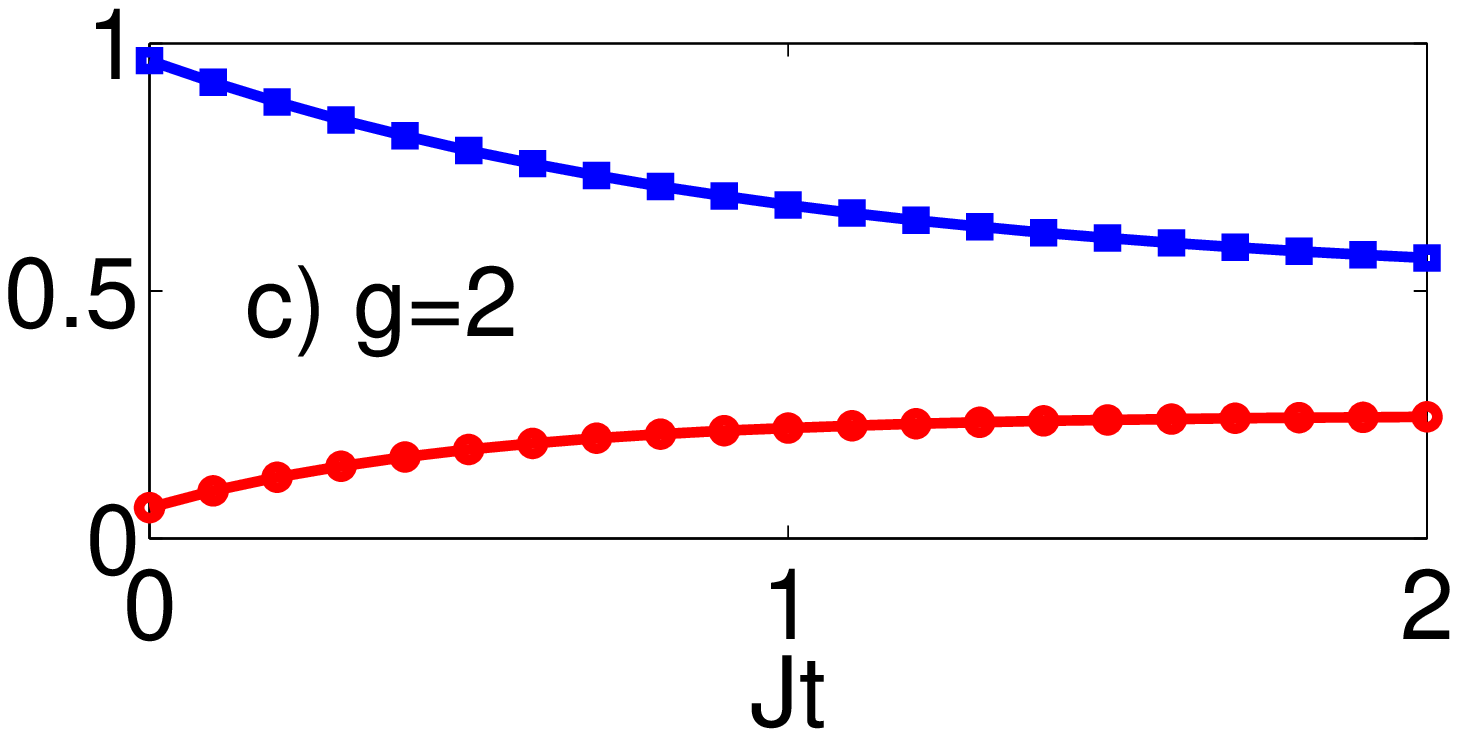,width=0.8\linewidth}
\caption{Thermalization: Mean (blue squares) $\bar{m}/N$ and variance $\sigma^2/N$ (red circles) for
increasing coupling time $t$ ($\gamma=\gamma_0=1$).  a)
$g=0$, b) $g=1$ and c) $g=2$.} \label{heatbath2}
\end{figure}

In fig. \ref{heatbath2}, we plot the mean and variance as a
function of time $t$ for a system coupled to a heat bath at
very high temperature $k_B T/J=100$. Here, the transverse field $g$ is fixed.
For no transverse field $g=0$ (fig. \ref{heatbath2} a ), both the
mean and the variance are constant as the coupling increases. In the critical point  $g=1$ (fig. \ref{heatbath2} b ), the
variance is constant and the mean decreases as the coupling time
increases. For high transverse field $g=2$ (fig. \ref{heatbath2}
c), the mean decreases until reaching the value of $0.5N$ and the
variance increases up to the value $0.25N$.

\subsection{Local representation of the coupling} \label{sec-local}
The master equation Eq. (\ref{eq-master-heat}) that we use to
describe thermalization shows two aspects. On the one hand, it is
physical to describe the coupling to the bath in terms of an
exchange of quasiparticles $\hat d_k$, because the Hamiltonian
Eq.(\ref{eq-ham1}) conserves the number of quasiparticle excitations. On the
other hand, it may look non-physical because the exchange between
the system and the bath is non-local. The aim of this section is
to show that the master equation can be rewritten in terms of
local fermions $\hat c_l$ and in principle it could be realized using
reservoir designs \cite{env}.

\begin{figure}
\centering
\epsfig{file= 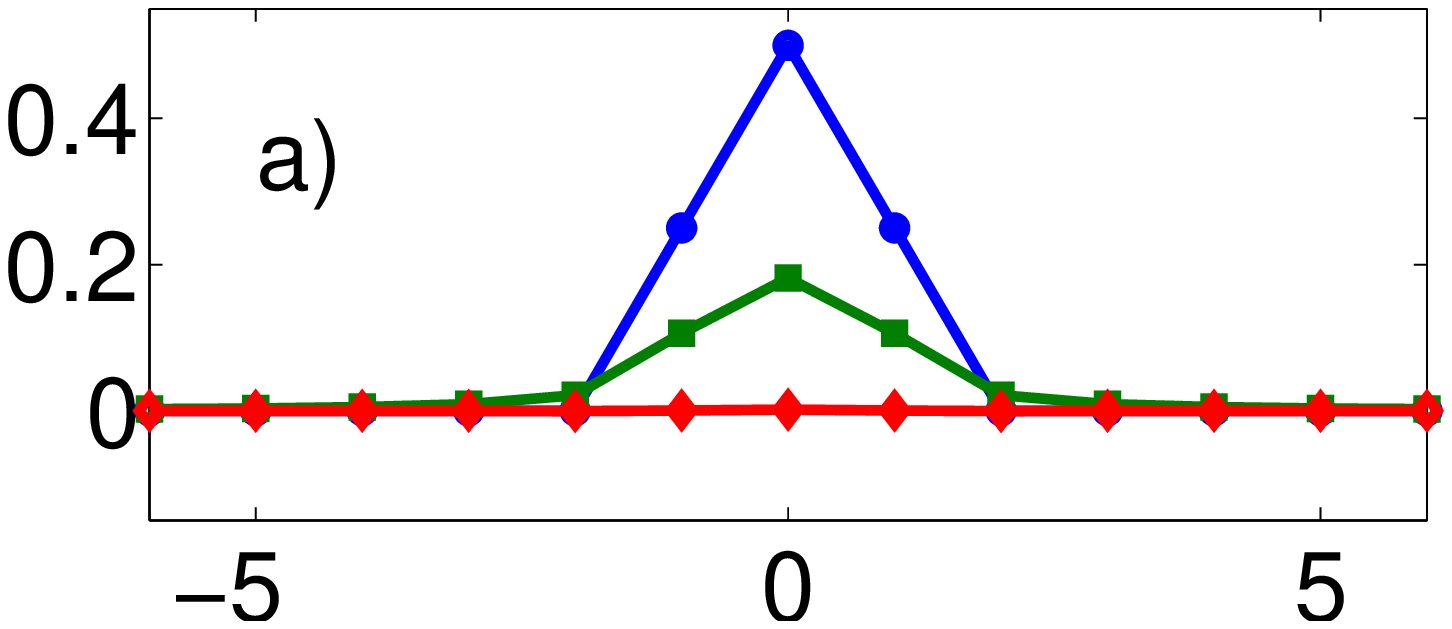,width=0.8\linewidth}  \epsfig{file=
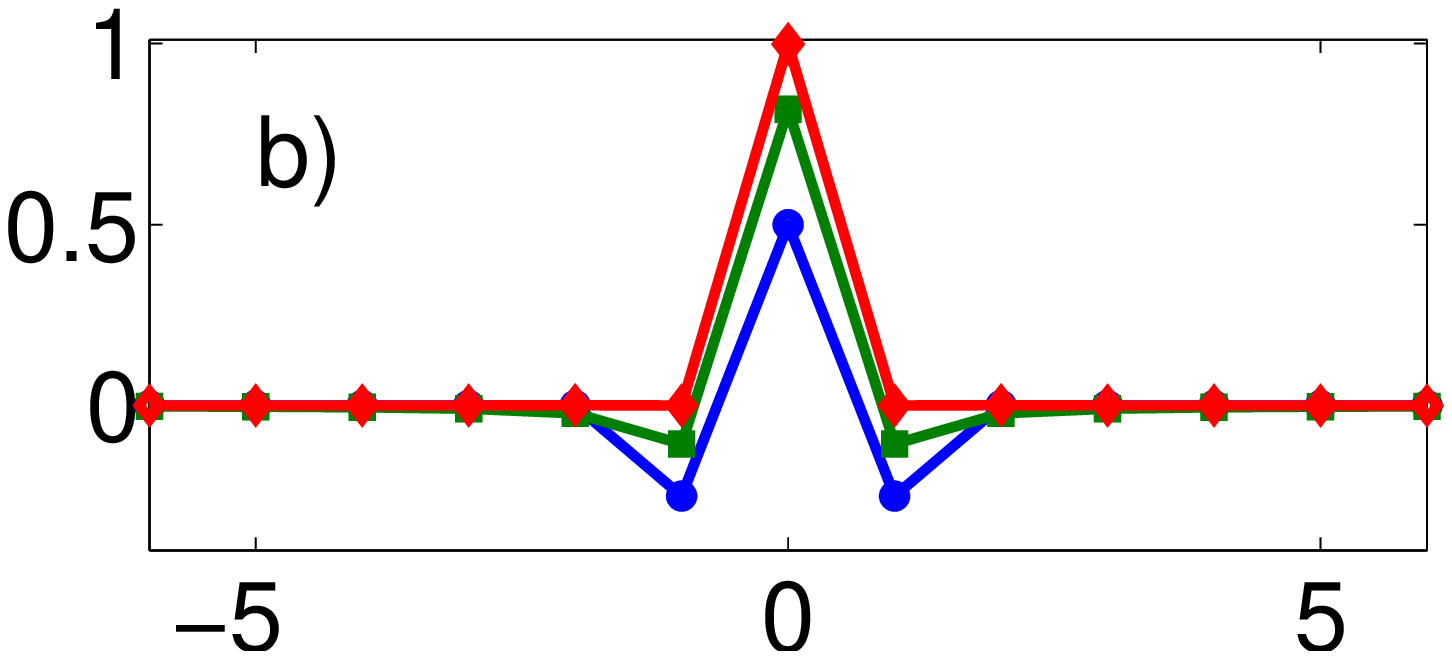,width=0.8\linewidth}  \epsfig{file=
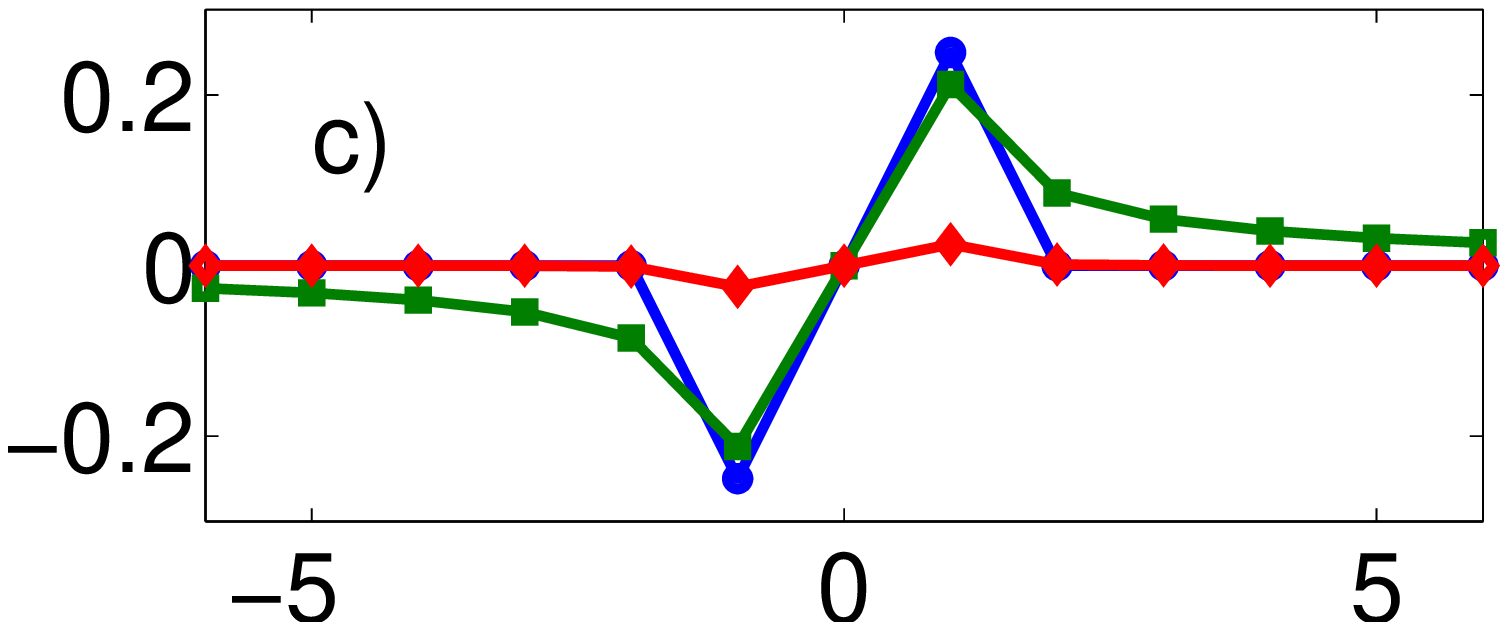,width=0.8\linewidth}
\caption{Thermalization: $F_u$ a), $F_v$ b) and $F_{uv}$ c)  as a function of distance between sites ($l-m$) for $\gamma=1$
and $g=0$ (blue line/circles), $g=1$ (green line/squares) and $g=10$
(red line/diamands). } \label{fs1}
\end{figure}

\begin{figure}
\centering
\epsfig{file= 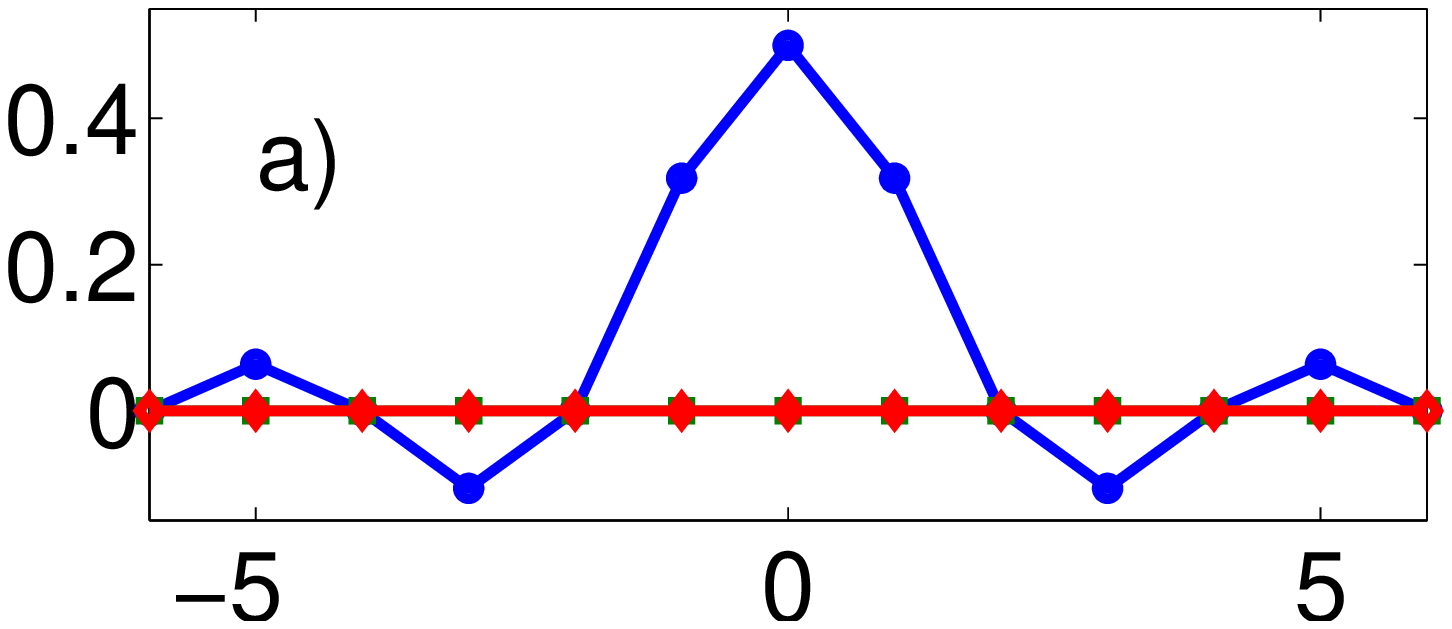,width=0.8\linewidth}  \epsfig{file=
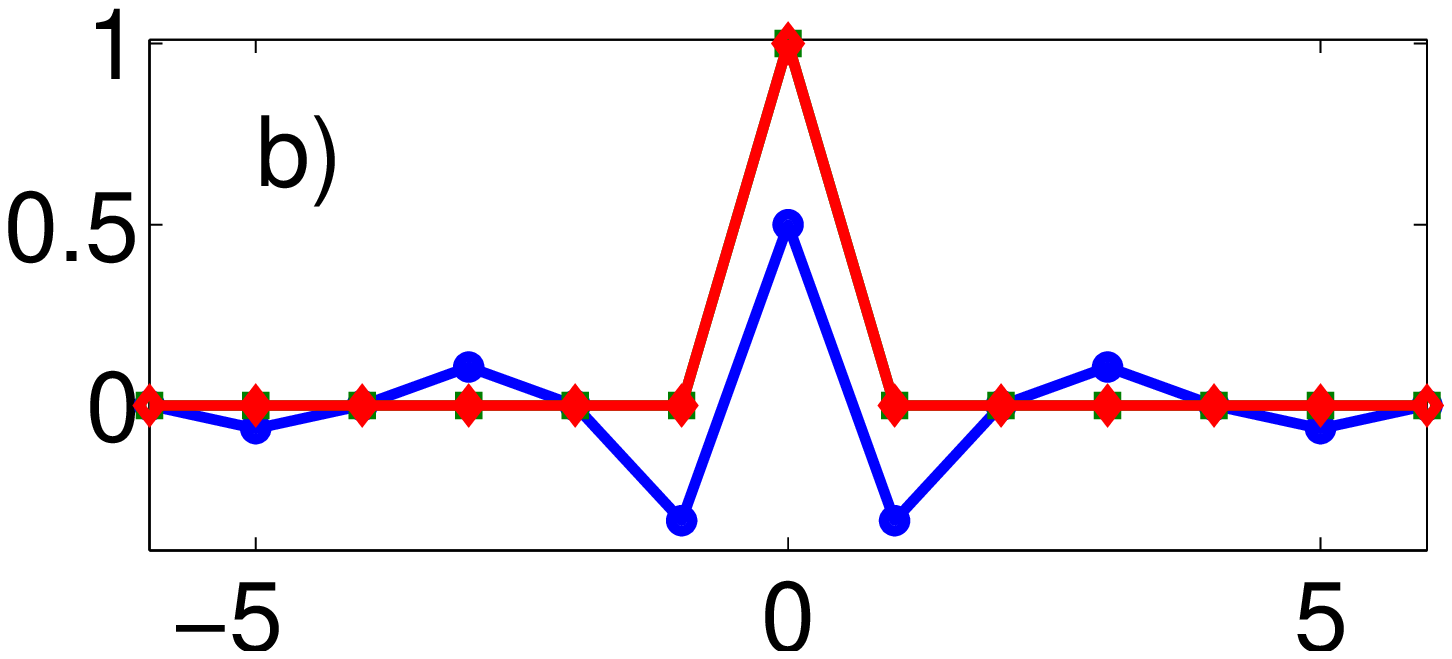,width=0.8\linewidth}  \epsfig{file=
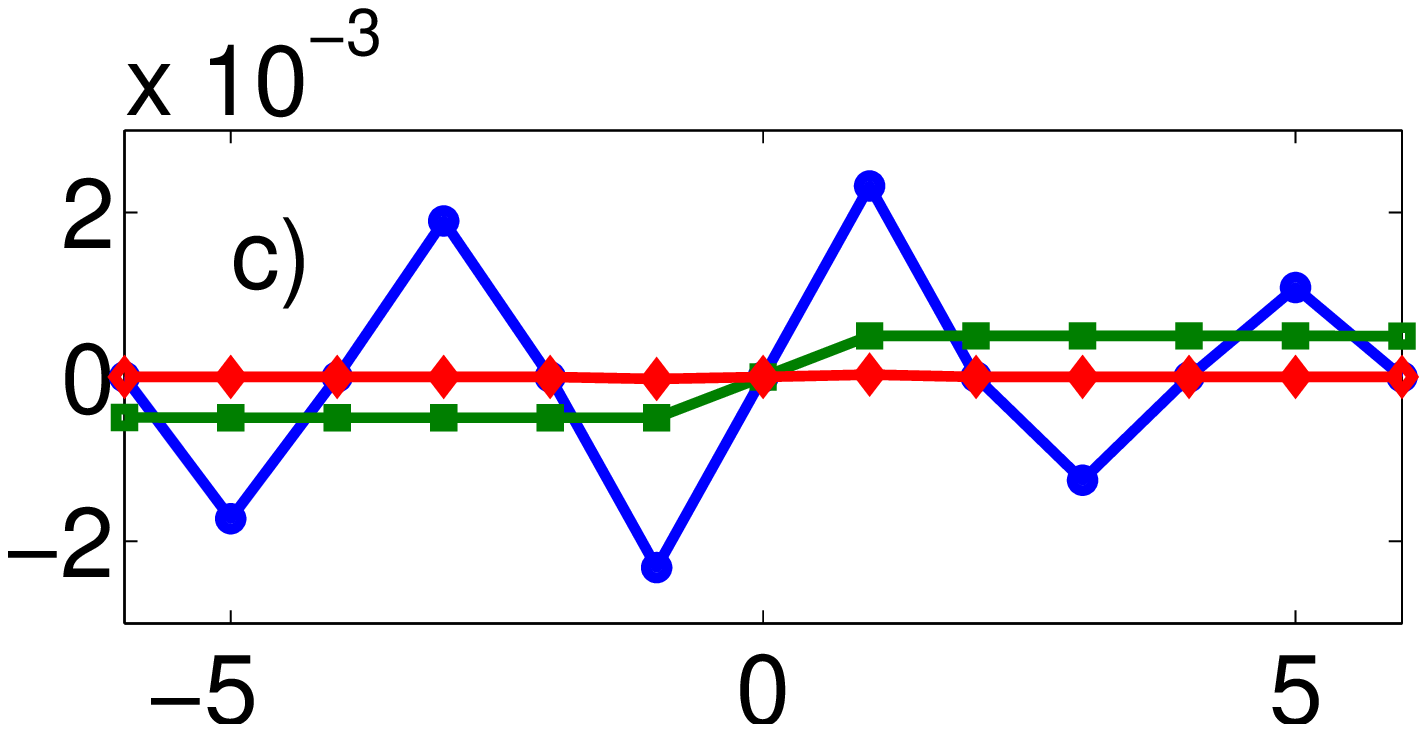,width=0.8\linewidth}
\caption{Thermalization: $F_u$ a), $F_v$ b) and $F_{uv}$ c) as a function of distance between sites $l-m$ for
$\gamma=0.01$ and $g=0$ (blue line/circles), $g=1$ (green
line/squares) and $g=10$ (red line/diamands). } \label{fs2}
\end{figure}

At high temperatures and in the absence of a transverse field ($g=0$)
at any temperature, the number of excitations $N_k^d$ in the bath
is constant with $k$. In this case, the master equation
(\ref{eq-master-heat}) in terms of the local operators $\hat c_l$
reads
\begin{widetext}
\begin{eqnarray}
\frac{d}{dt}\rho(t)=\nonumber\\
\gamma_0(\frac{N_d}{N}+1)\sum_{l,m}[F_u(l-m) \hat c_l \rho \hat
c_m^\dag+F_v(l-m) \hat c_l^\dag\rho \hat c_m-F_{uv}(l-m)(\hat
c_l^\dag\rho \hat c_m^\dag- \hat c_l\rho
\hat c_m)\nonumber\\
-\frac{1}{2}(F_u(l-m) \hat c_l^\dag \hat c_m\rho+F_v(l-m) \hat c_l
\hat c_m^\dag\rho-F_{uv}(l-m)( \hat c_l^\dag \hat c_m^\dag\rho
- \hat c_l  \hat c_m\rho))\nonumber\\
-\frac{1}{2}(F_u(l-m)\rho \hat c_l^\dag \hat c_m+F_v(l-m)\rho \hat
c_l \hat c_m^\dag-F_{uv}(l-m)(\rho \hat c_l^\dag \hat c_m^\dag
-\rho \hat c_l \hat c_m))]\nonumber\\
+\gamma_0\sum_k \frac{N_d}{N}\sum_{l,m}[F_u(l-m) \hat c_l^\dag\rho
\hat c_m+F_v(l-m) \hat c_l\rho \hat c_m^\dag-F_{uv}(l-m)( \hat
c_l^\dag\rho \hat c_m^\dag-\hat c_l\rho
\hat c_m)\nonumber\\
-\frac{1}{2}(F_u(l-m) \hat c_l \hat c_m^\dag\rho+F_v(l-m) \hat
c_l^\dag \hat c_m\rho-F_{uv}(l-m)( \hat c_l^\dag \hat c_m^\dag\rho
-\hat c_l \hat c_m\rho))\nonumber\\
-\frac{1}{2}(F_u(l-m)\rho \hat c_l \hat c_m^\dag+F_v(l-m)\rho \hat
c_l^\dag \hat c_m-F_{uv}(l-m)(\rho \hat c_l^\dag \hat c_m^\dag
-\rho \hat c_l \hat c_m))],\label{eq-master-local}
\end{eqnarray}
\end{widetext}
where we define the functions
\begin{eqnarray}
F_u(l-m)=\frac{1}{N}\sum_k  u_k^2 e^{i\Phi_k(l-m)}\\
F_v(l-m)=\frac{1}{N}\ \sum_k v_k^2e^{i\Phi_k(l-m)}\\
F_{uv}(l-m)=\frac{i}{N}\sum_k u_kv_k e^{i\Phi_k(l-m)},\\
\end{eqnarray}
which depend on the distance $l-m$ between two sites $l$ and $m$
and are related to the correlation length of the quasiparticles
and the pairs. In fig. \ref{fs1} and fig. \ref{fs2}, we study the
behavior of the functions $F_u$, $F_v$ and $F_{uv}$ as the
distance between the sites increases. We plot $F_u$, $F_v$ and
$\frac{1}{i}F_{uv}$ for different values of $g$ and $\gamma/J$ and
show that the functions $F_u$, $F_v$ have their maximum at zero
distance and decay rapidly as the distance increases. The function
$F_{uv}$, which corresponds to the pair correlations, has its
maximum at the nearest neighbor term $l-m=1$. We observe that for
large transverse field $g \gg 1$, and $\gamma/J \rightarrow 0$,
the only non-zero term corresponds to $F_v(0)=1$. In this case,
the XY model behaves like a free fermi gas and the master equation
(\ref{eq-master-heat}) reduces to
\begin{eqnarray}
\frac{d}{dt}\rho(t)=\nonumber\\
\gamma_0 (N_d/N+1)\sum_{l}[\hat c_l^\dag\rho \hat c_l-
\frac{1}{2}\hat c_l \hat c_l^\dag\rho
- \frac{1}{2} \rho \hat c_l \hat c_l^\dag]\nonumber\\
+\gamma_0 N_d/N\sum_{l}[ \hat c_l\rho \hat c_l^\dag - \frac{1}{2}
\hat c_l^\dag \hat c_l\rho - \frac{1}{2} \rho \hat c_l^\dag \hat
c_l].
\end{eqnarray}
Note that for these parameters, the quasiparticles $\hat d_k
\rightarrow \hat c_k^\dagger$. Thus at high $T$ and high
transverse field $g$ the bath and the system exchange fermionic
particles.

Another interesting limit occurs at any $T$ when $g\rightarrow 0$
and $\gamma/J=1$. We see in figure \ref{fs1} that in this case,
the functions $F_u$, $F_v$ are of order $0.5$ for the same site
and $F_u$, $F_v$ and $F_{uv}$ are of the order of $\pm 0.25$ for
neighboring sites. The master equation Eq. (\ref{eq-master-local})
has contributions from exchange of on-site fermions and an
additional term that corresponds to neighboring particles. Also,
there is exchange of not only on-site particles and holes but also
fermionic pairs. This is expected as in this regime $g\ll
\gamma,J$  the pair creation dominates in Hamiltonian Eq.
(\ref{eq-ham1}).

For low temperatures and at $g\neq0$, the number of quasiparticles
$N_k^d$ is not constant with $k$ and the master equation cannot be
written in the form Eq. (\ref{eq-master-local}). However, as $N_k^d$
is small for low temperatures, the non-local terms are negligible
and the equation as a whole remains local.
\section{Conclusions}
\label{sec-conc} We have studied the effect of temperature in the
counting distribution of a strongly correlated fermionic system
which can be mapped to the quantum XY spin model with a transverse field.
Thermal fluctuations induce pair breaking in the superfluid fermionic system.  We show that this is reflected in the number distribution function which becomes non-zero for odd number of particles for a temperature proportional to the pair formation strength.  Also, thermal fluctuations reduce the quantum phase transition into a crossover between different regions of the phase diagram.
We have found that at low temperatures, the mean and variance of the
counting distribution reflect the critical behavior at the
crossover between different phase regimes. This effect is obscured with increasing temperature and when the temperature is
comparable to the eigenergies of the system, the critical
behavior is blurred from the cumulants of the counting
distribution.

Furthermore, we have shown that the number distribution functions can be used to monitor the quantum dynamics of the system. We have studied the thermalization of the
system initially at zero temperature when it is coupled to a heat bath at finite temperature. This process is analogous to a temperature quench.
The  temperature determines the number of delocalized excitations
in the system at equilibrium. For high temperatures and high
transverse fields, the exchange of excitations between system and
bath can be mapped into the exchange of local fermions. For zero
transverse field, we have shown that the exchange of local
excitations corresponds to the exchange of local particles and
nearest neighbor pairs. We have assumed that the counting process
occurs at a different time scale, much faster than the exchange of
excitations between the system and the bath. We have shown that
the mean and variance of the counting distribution can be used to
map the thermalization process.

\section*{Appendix: Counting of constant fields}
The counting formula in Eq. (\ref{eq-p}) has been derived in
different ways \cite{Glauber-seminal,Mandel,
Kelley,lax,mollow,scully,selloni,srinivas}. Here, we review a
derivation (see e.g. \cite{Ueda1990}) by modeling the absorption
of the particles at the detector with the master equation
\begin{equation}
\dot{\rho}=\varepsilon  \hat a\rho \hat
a^\dag-\frac{\varepsilon}{2}\hat a^\dag \hat a
\rho-\frac{\varepsilon}{2}\rho \hat a^\dag \hat a,\label{master}
\end{equation}
where $a^\dag$ and $a$ are the creation and annihilation operator
of the particle to be counted. Performing a rotation of the
density matrix, $\rho(t)=e^{-\frac{\varepsilon}{2}t \hat a^\dag
\hat a}\tilde{\rho}(t)e^{-\frac{\varepsilon}{2}t\hat a^\dag \hat
a}$, and using the relation
\begin{equation}
e^{\gamma A}Be^{-\gamma
A}=B+\gamma[A,B]+\frac{\gamma^2}{2!}[A,[A,B]]+...,
\end{equation}
we obtain
\begin{equation}
\dot{\tilde{\rho}}(t)=\varepsilon \hat a
e^{\frac{-\varepsilon}{2}t}\tilde{\rho}\hat a^\dag
e^{\frac{-\varepsilon}{2}t}=\varepsilon e^{-\varepsilon
t}\hat a\tilde{\rho}\hat a^\dag.
\end{equation}
This equation can be solved using perturbation theory
\begin{eqnarray}
\tilde{\rho}(t)=\tilde{\rho}(0)+\int_0^t\varepsilon e^{-\varepsilon
t'}\hat a\tilde{\rho}(t')\hat a^\dag.\nonumber\\
\end{eqnarray}
Transforming back the rotation we obtain
\begin{equation}
\rho(t)=e^{-\frac{\varepsilon}{2}t \hat a^\dag
\hat a}(\tilde{\rho}(0)+\int_0^t\varepsilon e^{-\varepsilon t'}
\hat a\tilde{\rho}(0)\hat a^\dag+...)e^{-\frac{\varepsilon}{2}t \hat a^\dag \hat a}.
\end{equation}
Using the cyclic properties of the trace, the probability $p_m(t)$
of counting $m$ particles can be written as
\begin{equation}
p_m(t)=\hbox{Tr} [\rho (0) a^{\dag m}\frac{(\int_0^tdt'\varepsilon
e^{-\varepsilon t'})^m}{m!}e^{-\varepsilon t a^\dag a}a^m].
\end{equation}
This is equal to the normally ordered expression
\begin{equation}
p_m(t)=\langle:(1-e^{-\varepsilon t})^m\frac{(\hat a^\dag
\hat a)^m}{m!}e^{-(1-e^{-\varepsilon t})\hat a^\dag \hat a}:\rangle,
\end{equation}
which holds because
\begin{eqnarray}
&:(\hat a^\dag
\hat a)^m e^{-(1-e^{-\varepsilon t})\hat a^\dag \hat a}:=\hat a^{\dag m}:  e^{-(1-e^{-\varepsilon t})\hat a^\dag \hat a}:\hat a^ {m}\nonumber \\
&=\hat a^{\dag m} e^{-\varepsilon t \hat a^\dag \hat a} \hat a^ {m}.
\end{eqnarray}
We can thus use the generating function formalism in
Eq.(\ref{eq-Q-a}) with $\kappa=\int_0^\tau dt' \varepsilon
e^{-\varepsilon t'}=1-\exp{(-\varepsilon \tau)}$, where $\tau$ is
the aperture time of the detector.

\begin{acknowledgments}
We acknowledge financial support from the Spanish MINCIN project
FIS2008-00784 (TOQATA), Consolider Ingenio 2010 QOIT, EU STREP
project NAMEQUAM, ERC Advanced Grant QUAGATUA, the Ministry of
Education of the Generalitat de Catalunya, and from the Humboldt
Foundation. Mirta Rodr\'iguez is grateful to the MICINN of Spain for a Ram\'on y Cajal contract.
\end{acknowledgments}

\end{document}